\documentclass{aa}

\usepackage{graphicx}
\usepackage{txfonts}
\begin{document}

   \title{The growth history of local M\,33-mass bulgeless spiral galaxies}

%   \subtitle{I. Overviewing the $\kappa$-mechanism}

   \author{Xiaoyu~Kang
          \inst{1,2,3}
          \and
          Rolf-Peter~Kudritzki\inst{4,5}
          \and
          Fenghui~Zhang\inst{1,2,3}
          }

   \institute{Yunnan Observatories, Chinese Academy of Sciences, 396 Yangfangwang,
   Guandu District, Kunming, 650216, P.R. China\\
              \email{kxyysl@ynao.ac.cn}
         \and
             Key Laboratory for the Structure and Evolution of Celestial Objects, Chinese Academy of Sciences, 396 Yangfangwang, Guandu District, Kunming, 650216, P. R. China
         \and International Centre of Supernovae, Yunnan Key Laboratory, Kunming 650216, P. R. China
         \and LMU M$\rm \ddot{u}$nchen, Universit$\rm \ddot{a}$tssternwarte, Scheinerstr. 1, 81679 M$\rm \ddot{u}$nchen, Germany
         \and Institute for Astronomy, University of Hawaii, 2680 Woodlawn Drive, Honolulu, HI96822, USA
             }

   \date{Received August 9, 2023; accepted September 14, 2023}

 \abstract
 {NGC\,7793, NGC\,300, M33 and NGC\,2403 are four nearby undisturbed and
bulgeless low-mass spiral galaxies with similar morphology and stellar mass.
They are ideal laboratories to study disc formation scenarios and stellar mass
growth histories. We construct a simple chemical evolution model by assuming that
discs grow gradually with continuous metal-free gas infall and metal-enriched gas
outflow. By means of the classical $\chi^{2}$ methodology, applied to the model
predictions, the best combination of free parameters capable of reproducing the
corresponding present-day observations is determined, i.e. the radial
dependence of the infall timescale $\tau=0.1r/{R_{\rm d}}+3.4\,{\rm Gyr}$
($R_{\rm d}$ is the disc scale-length) and the
gas outflow efficiency $b_{\rm out}=0.2$. The model results are in excellent
agreement with the general predictions of the inside-out growth scenario
for the evolution of spiral galaxies. About 80\% of the stellar mass of
NGC\,7793 is assembled within the last 8\,Gyr and 40\% within the last
4\,Gyr. By comparing the best-fitting model results of the
three other galaxies we obtain similar results, 72\% (NGC\,300), 66\% (NGC\,2403) and 79\% (M33)
stellar mass were assembled within the past $\sim\rm 8\,Gyr$ (i.e. $z\,=\,1$).
These four disc galaxies simultaneously increase their sizes and stellar masses
as time goes by and they grow in size at $\sim\,0.30$ times the rate at which
they grow in mass. The scale-lengths of these four discs are now 20\% -- 25\%
larger than at $z\,=\,1$.
Our best-fitting model predicted the stellar mass-metallicity relation and the metallicity
gradients, constrained by the observed metallicities from HII-regions emission line analysis,
agree well with the observations measured from individual massive red and blue supergiant stars
and population synthesis of SDSS galaxies.}

%  \abstract
   {
   }
   {}
   {}
   {}
   {}


   \maketitle
%
%-------------------------------------------------------------------

\section{Introduction}
\label{sec:intro}

Understanding how galaxies assemble is crucial to understanding
the overall picture of galaxy evolution. Spiral galaxies have a
major role in the universe today, with disc galaxies dominating
star formation in the present epoch \citep{Brinchmann2004, Williams2011}
and keeping forming stars at a sustained rate throughout their
evolution \citep{Aumer2009, Fraternali2012, Pezzulli2015}.
At the same time, the structure and star formation history (SFH) of disc
galaxies are tightly linked with stellar mass \citep{Casasola2017},
and galaxies with stellar masses near or below Milky Way--mass
(${\rm log}(M_{\rm MW,\ast}/{\rm M}_{\odot})\,\sim\,10.7$)
assemble their mass mainly from in situ star formation, not from
mergers \citep{Leitner2011, Qu2017, Behroozi2019, Pan2019}.
Furthermore, recent studies about deep surveys detecting galaxy
masses (${\rm log}(M_{\ast}/{\rm M}_{\odot})\,<\,10$) have shown that
very large populations low-mass galaxies exist at all redshifts and
galaxy stellar mass functions exhibit a slightly steeper low-mass end
\citep{Baldry2008, Baldry2012, Bauer2013, Kelvin2014},
emphasizing a necessity to understand the growth of low-mass galaxies,
the most populous in the universe.

Most of spiral galaxies are difficult to study in detail due
to the complexities of their bulges \citep{Byun1995} or
mergers \citep{Barnes1992}. While bulgeless discs
appear to be excellent laboratories to study the secular evolution
of galactic discs, most of them are too distant to study in great detail.
Fortunately, in the nearby Universe ($D\,<\,4\,\rm Mpc$), there are
four undisturbed and bulgeless (pure-disc) galaxies with similar morphology
and stellar mass (${\rm log}(M_{\ast}/{\rm M}_{\odot})\,<\,10$):
NGC\,7793 \citep{Sacchi2019}, NGC\,300 \citep{Gogarten2010},
M33 \citep{Williams2009} and NGC\,2403 \citep{Williams2013}.
The wealth of available data including both local and global data,
such as gas mass, star formation rate (SFR), stellar mass
\citep[e.g. ][]{Leroy2008}, gas-phase metallicity \citep[e.g. ][]{Pilyugin2014}
and specific SFR \citep[sSFR,][]{MM2007, Smith2021},
for these four low-mass spiral galaxies can be used to constrain their theoretical
models, therefore we can use their theoretical results to compare the detailed
evolution of these discs. Such comparative studies offer an ideal opportunity to
search for clues to the processes that drive the galaxy disc growth histories,
for instance, inside-out formation scenario, namely that
the inner parts of the galaxy disc forms faster than the outer ones
\citep[e.g. ][]{Larson1976, Matteucci1989, Chiappini2001, Grisoni2018, Frankel2019, Spitoni2021a}.

On the other hand, observations in local star-forming galaxies have
revealed the existence of a tight relation between stellar mass and gas
metallicity: the stellar mass-metallicity relation (MZR), with the more massive galaxies
being more metal enriched
\citep[e.g. ][among many others]{Tremonti2004, Kewley2008, Andrews2013, Zahid2014}.
As for the stellar metallicity, the MZR is also found in star-forming galaxies
\citep[e.g. ][]{Gallazzi2005, Panter2008, Thomas2010, Zahid2017, Sextl2023}.
\citet{Bresolin2022} updated the MZR based on metallicity studies of stars in
resolved star-forming galaxies. The MZR seems to be a Rosetta stone for understanding
the formation and evolution of galaxies.

In addition to the global metallicities, the radial gradient
in metallicity is also important to understand the formation history
of galaxies. After the pioneering work by \citet{Aller1942}, the radial
metallicity gradient has been studied for a long time.
Based on long-slit spectroscopy and Integral field unit (IFU) spectroscopy,
most spiral galaxies in the local Universe exhibit negative metallicity
gradients in gas and stars within their optical radius, that is,
the centre of a galaxy has a higher metallicity than the outskirts
\citep[e.g. ][]{Zaritsky1994, Moustakas2010, Sanchez2014, Pilyugin2014, Pilyugin2019, Gonzalez2015, Ho2015, Zheng2017, Goddard2017, Bresolin2019, Bresolin2022}.

At the same time, spiral galaxies that show no clear evidence of
an interaction present a common metallicity gradient, if normalized to
an appropriate scale-length, such as the disc effective radius
\citep[$R_{\rm e}$, e.g. ][]{Sanchez2014}, the disc scale-length
\citep[$R_{\rm d}$, e.g. ][]{Garnett1997} and the isophotal radius
\citep[$R_{25}$, e.g. ][]{Zaritsky1994, Pilyugin2014, Pilyugin2019, Ho2015, Bresolin2019},
and the slope is independent of galaxy properties, such as stellar mass,
absolute magnitude, morphology, or whether a bar is present or not.

The simple chemical evolution model has been widely used in
studies of the chemical evolution and SFHs of nearby disc galaxies
\citep[e.g. ][]{Tinsley1980, Chang1999, Chang2012, BP2000, Chiappini2001, Molla2005, Kubryk2015, Kang2017, Bresolin2019},
making great progress in our understanding the formation and evolution
of disc galaxies.
Meanwhile, the SFH is a vital aspect of the formation and evolution of any disc
galaxies, since it dominates its resulting stellar disc structure \citep{Frankel2019}.
Thus, in this work, the simple chemical evolution model is adopted
to study the chemical evolution and SFHs of M\,33-mass bulgeless spiral galaxies.

The main aim of this paper is to investigate the disc
formation scenarios and stellar mass growth histories of M\,33-mass
bulgeless spiral galaxies, as well as to study whether or not the model
predicted results of these four galaxies can explain the MZR
and the observed metallicity gradients.
The chemical evolution and SFHs of M33, NGC\,300 and NGC\,2403
have been studied in our previous work by using the simple chemical evolution model
\citep{Kang2012, Kang2016, Kang2017}.
Therefore, before making comparative studies of these four galaxies,
we should construct a simple chemical evolution model to
explore the chemical evolution and SFH of NGC\,7793 first.

The outline of this paper is organized as follows. Section\,\ref{sec:model}
describes the main ingredients of the chemical evolution model. The observed data
including radial profiles and global properties of atomic hydrogen (HI) gas, SFR, sSFR and
gas-phase metallicity constraints of NGC\,7793 are presented in Sect.\,\ref{sec:observe}.
In Sect.\,\ref{sec:result}, we investigate the chemical evolution and SFH of
NGC\,7793 by using the simple chemical evolution model, and provide the disc
formation scenarios and stellar mass growth histories of M\,33-mass spiral
galaxies by comparing the evolution and SFH of NGC\,7793 with those of M\,33,
NGC\,300 and NGC\,2403.
Conclusions are given in Sect.\,\ref{sec:sum}.

%--------------------------------------------------------------------

\section{Model}
\label{sec:model}

Similar to our previous work \citep{Chang1999, Kang2012, Kang2016, Kang2017},
a star-forming galaxy disc is assumed to gradually build up due to continuous
infall of metal-free gas ($X\,=\,0.7571, Y_{\rm p}\,=\,0.2429, Z\,=\,0$) from its
halo, and it is composed of a set of independently evolved concentric rings, in
the sense that no radial mass flows is allowed in the model.
In fact, such flows can take place in real galaxies as a result of the presence of the
bars \citep{Sellwood1993}, redistribution of angular momentum owing to viscosity
\citep{Ferguson2001} and radial stellar migration \citep{Rovskar2008}.
In particular, the age-metallicity relationship in the solar neighbourhood of
our Galaxy \citep[e.g. ][]{Edvardsson1993, Haywood2008, Schonrich2009, Minchev2010, Feuillet2019, Xiang2022}
and the observed U-shaped colour profiles in galaxies \citep{Azzollini2008, Bakos2008}
put strong constraints on the presence of radial migration
\citep[e.g. ][]{Schonrich2009, Minchev2012, Kubryk2013, Spitoni2015, Frankel2018, Buck2020, Vincenzo2020, Chen2023}.
Moreover, the metal-enriched gas outflows are considered in the model, since both observations
\citep{Garnett2002, Tremonti2004} and theoretical models \citep{Erb2008, Finlator2008, Spitoni2020}
showed that the gas outflow process plays a key role during the chemical evolution
of low-mass galaxies with stellar mass ${\rm log}(M_{\ast}/{\rm M}_{\odot})\,<\,10.5$.
Finally, the model does not include the bulge nor does it differentiate between
the thin and thick disc \citep[see e.g. ][]{Grisoni2017, Spitoni2019}.
Despite the simplifying assumption of independently evolving rings, the
model has been successfully applied to investigate the formation and
evolution of the Milky Way disc and other nearby disc galaxies
\citep{Chang1999, BP2000, MM2011, Kang2012, Kang2016, Kang2017, Bouquin2018, Bresolin2019}.

Thus, main ingredients of the model are including infalls of metal-free gas, star formation,
metal production via stellar evolution, stellar mass return and outflows of
metal-enriched gas. The details of these ingredients are presented in the following.
The instantaneous recycling approximation \citep[IRA,][]{Tinsley1980} is adopted in the model by
assuming that the gas return from stars to the interstellar medium (ISM) happens on a short timescale
compared with galactic evolution, and the gas is well mixed with stellar ejecta.
Although IRA is a strong assumption in chemical evolution modelling, it still represents
a good approximation for the abundance of chemical elements produced by massive stars with short
lifetimes. The best example of such a chemical element is given by oxygen, which also represents
the best proxy for the global metallicity of the galaxy ISM, since it is the most abundant heavy
element by mass. On the other hand, the ISM evolution of chemical elements produced by stars with
long lifetimes cannot be followed by chemical evolution models working with the IRA assumption.
Examples of such chemical elements are given by nitrogen, carbon and iron. To consider the stellar
lifetimes with a high level of detail, numerical chemical evolution models, in which the IRA is
relaxed, should be adopted \citep[see][]{Matteucci2012, Vincenzo2016}.

The evolution in each ring can be described
by the following differential equations from \citet{Tinsley1980}:
\begin{equation}
%\begin{array}{l}
\frac{{\rm d}[\Sigma_{\rm tot}(r,t)]}{{\rm d}t}\,=\,f_{\rm{in}}(r,t)-f_{\rm{out}}(r,t),\\
\label{eq:tot}
\end{equation}
\begin{equation}
\frac{{\rm d}[\Sigma_{\rm gas}(r,t)]}{{\rm d}t}\,=\,-(1-R)\Psi(r,t)+f_{\rm{in}}(r,t)-f_{\rm{out}}(r,t),\\
\label{eq:gas}
\end{equation}
\begin{eqnarray}
\frac{{\rm d}[Z(r,t)\Sigma_{\rm gas}(r,t)]}{{\rm d}t}\,=\,y(1-R)\Psi(r,t)-Z(r,t)(1-R)\Psi(r,t) \nonumber\\
+Z_{\rm{in}}f_{\rm{in}}(r,t)-Z_{\rm{out}}(r,t)f_{\rm{out}}(r,t),
\label{eq:metallicity}
\end{eqnarray}
where $\Sigma_{\rm tot}(r,t)$ and $\Sigma_{\rm gas}(r,t)$ are the total
(star + gas) and gas mass surface density in the ring centered at
galactocentric distance $r$ at evolution time $t$,  respectively;
$\Psi(r,t)$ and $Z(r,t)$ are the SFR and
metallicity in the corresponding place and time, respectively.
$f_{\rm{in}}(r,t)$ and $f_{\rm{out}}(r,t)$ are the gas infall rate
and the gas outflow rate in the corresponding place and time, respectively.
$R$ is the return fraction and $y$ is the nuclear synthesis yield.
Both $R$ and $y$ are functions of the initial mass function (IMF) and weakly
depend on metallicity and time \citep{Vincenzo2016}. Neither metallicity nor time
dependence of $R$ and $y$ will be further considered in our model, since the
oxygen yield is independent of metallicity and the stellar lifetimes under our
assumptions.
Furthermore, galactic chemical enrichment depends significantly on the IMF
\citep{Vincenzo2016, Goswami2021} and the \citet{Kroupa1993} IMF is favoured in
describing the chemical evolution of spiral discs \citep{Romano2010, Vincenzo2016},
therefore the IMF of \citet{Kroupa1993} is adopted in this work.
The values of $R\,=\,0.289$ and $y\,=\,0.019$ are obtained by averaging the
values of $R$ and $y_{Z}$ over metallicity in Table\,2 of \citet{Vincenzo2016},
corresponding to the compilation of stellar yields of \citet{Romano2010}.
$Z_{\rm{in}}$ is the metallicity of the infalling gas and assumed to be metal-free,
i.e. $Z_{\rm{in}}=0$. $Z_{\rm{out}}(r,t)$ is the metallicity of the outflowing gas and
assumed to have the same metallicity as the ISM, i.e. $Z_{\rm{out}}(r,t)=Z(r,t)$
\citep{Chang2010, Belfiore2016, Kang2012, Kang2016, Kang2017, Kang2021}.

\begin{figure}
  \centering
  \includegraphics[angle=0,width=0.475\textwidth]{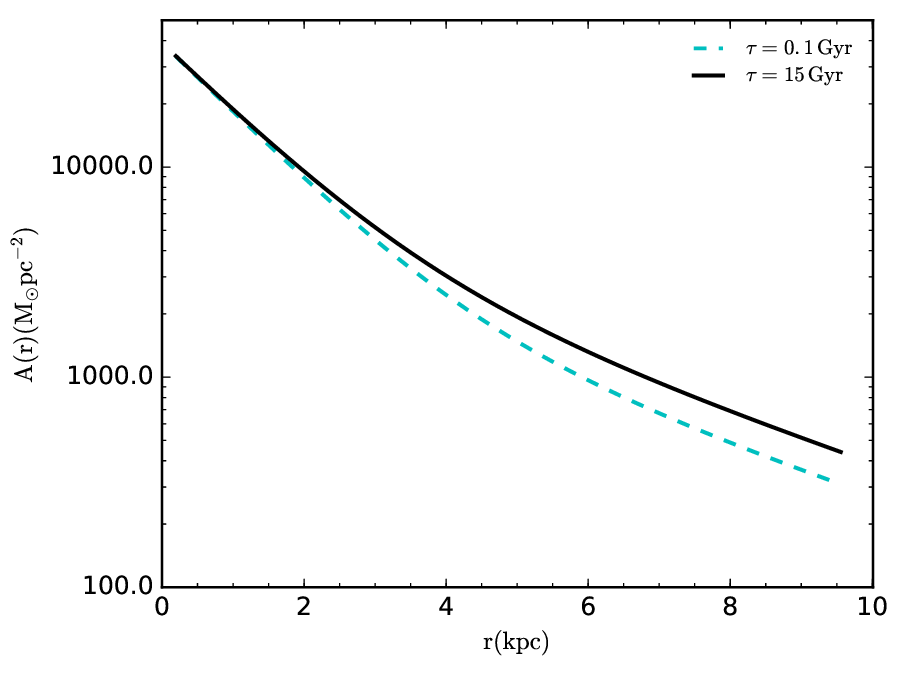}
    \caption{$A(r)$ function. The dashed and solid lines
    are corresponding to two limiting cases of $\tau\,=\,0.1\,\rm Gyr$
    and $\tau\,=\,15\,\rm Gyr$, respectively. Note that, for the convenience
    of comparison, the $A(r)$ values of the model adopting $\tau\,=\,15\,\rm Gyr$
    have been multiplied by 4000.
    }
  \label{Fig:A_r}
\end{figure}

The metal-free gas infall rate at each radius $r$ and time $t$,
$f_{\rm in}(r, t)$, in units of $\rm M_\odot\,pc^{-2}\,Gyr^{-1}$,
as a function of space and time, is adopted as in \citet{Kang2012,Kang2016,Kang2017}:
\begin{equation}
f_{\rm{in}}(r,t)=A(r)\cdot t\cdot e^{-t/\tau},
\label{eq:infall rate}
\end{equation}
where $\tau$ is the gas infall timescale, and it is a free parameter
in our model.
The range of the free parameter $\tau$ is from $0.1\,\rm Gyr$ to $15.0\,\rm Gyr$.
As shown in Fig. 1 of \citet{Kang2016}, the gas infall rate we adopted in this work
is low in the beginning and gradually increases with time. This gas infall rate reaches
the maximal value when $t\,=\,\tau$ and then falls. The case of $\tau\,=\,0.1\,\rm Gyr$
is corresponding to a time-declining gas infall rate that most of the cold gas has been
accreted to the disc in the early period of its history, while that of $\tau\,=\,15.0\,\rm Gyr$
represents a time-increasing gas infall rate that a large fraction of cold gas is now
falling onto the disc of NGC7793. How we obtain the best-fitting value of $\tau$ will
be presented in Sect.\,\ref{sec:result_r}.

The function $A(r)$ is determined iteratively so that the
present-day model predicted stellar mass surface density
$\Sigma_*(r,t_{\rm g})$ matches the observations
\citep{Kang2012,Kang2016,Kang2017}.  $\Sigma_*(r,t_{\rm g})$ is well
described by an exponential profile:
\begin{equation}
\Sigma_*(r,t_{\rm g})=\Sigma_*(0,t_{\rm g}){\rm exp}(-r/R_{\rm d}),
\label{eq:stellar_distrbution}
\end{equation}
where $R_{\rm d}$ is the present-day disc scale-length.
$\Sigma_*(0,t_{\rm g})$ is the present-day central stellar mass
surface density, and it can be easily obtained from
$\Sigma_*(0,t_{\rm g})=M_{*}/2\pi R_{\rm d}^{2}$. In this work, we adopt the
stellar mass and disc scale-length of NGC\,7793
as $M_{*}\,=\,3.162\times10^{9}~\rm M_{\odot}$
and $R_{\rm d}\,=\,1.3\,\rm kpc$, which are estimated
from Infrared Array Camera (IRAC) images at $3.6\,\mu m$
\citep{Leroy2008}. $t_{\rm g}$ is the cosmic age and we set
$t_{\rm g}=13.5\rm\,Gyr$ according to the standard flat cosmology.

In our calculation we adopt a value of the free parameter
$\tau$ and then start with a first estimate of $A(r)$ and numerically
solve Eqs.\,(\ref{eq:tot}) and (\ref{eq:gas}) adopting a SFR surface density
$\Psi(r,t)$ (see below). By comparing the model predicted $\Sigma_*(r,t_{\rm g})$ with its
observed value, we adjust the value of $A(r)$ and repeat
the calculation until the resulting $\Sigma_*(r,t_{\rm g})$
fits the observed radial distribution. Figure\,\ref{Fig:A_r} plots
the model predicted radial profile of $A(r)$, where the dashed and solid lines
are corresponding to two limiting cases of $\tau\,=\,0.1\,\rm Gyr$
and $\tau\,=\,15\,\rm Gyr$, respectively.
It should be emphasized that, for the convenience of comparison,
the $A(r)$ values of the model adopting $\tau\,=\,15\,\rm Gyr$
have been multiplied by 4000. As can be seen from Fig.\,\ref{Fig:A_r},
$A(r)$ varies strongly with $\tau$.
%\textbf{However, We do not treat $A(r)$ as a free parameter,
%since it is roughly fixed for given $\tau$.}

%\textbf{The natural emergence of the correlation between H 2 and star formation
%rate surface densities in galaxy simulations$\cdot\cdot\cdot$
%\citep{Lupi2018, Semenov2019}}

The SFR surface density $\Psi(r,t)$ (in units of
$\rm{M_{\odot}}\,{pc}^{-2}\,{Gyr}^{-1}$) describes how a disc
galaxy of a given cold gas mass and scale radius will form its stars.
\citet{Leroy2008} and \citet{Bigiel2008} have found that $\Psi(r,t)$ appears to
correlate strongly with the molecular hydrogen gas mass surface density
($\Sigma_{\rm H_2}(r,t)$) on sub-kpc scales, rather than the mass
surface density of atomic hydrogen gas ($\Sigma_{\rm HI}(r,t)$) or the
total gas ($\Sigma_{\rm tot}(r,t)\,=\,\Sigma_{\rm{H_{2}}}(r,t)+\Sigma_{\rm HI}(r,t)$).
They also found that in their sample of spiral galaxies
(including NGC\,7793)  the molecular hydrogen gas ($\rm H_2$) forms
stars at a roughly constant efficiency
at radii where it can be detected \citep{Leroy2008,Bigiel2008}.
$\Psi(r,t)$ proportional to $\Sigma_{\rm H_2}(r,t)$ has been used in disc models
by \citet{Lagos2011}, \citet{Kang2012,Kang2016,Kang2017} and \citet{Kubryk2015},
and it will be also adopted in this paper:
\begin{equation}
\Psi(r,t)=\Sigma_{\rm{H_2}}(r,t)/t_{\rm dep},
\label{eq:h2sfr}
\end{equation}
Where $t_{\rm dep}$ is the molecular gas depletion time, and its value
is adopted as $t_{\rm dep}\,=\,1.9\,\rm Gyr$ in this work \citep{Leroy2008,Bigiel2008}.
 In order to split $\Sigma_{\rm{gas}}(r,t)$ into $\Sigma_{\rm{H_{2}}}(r,t)$
and $\Sigma_{\rm HI}(r,t)$ in our galaxy evolution model, we adopt the semi-emi-empirical
prescription by \citet{Blitz2006} and \citet{Leroy2008} for the molecular-to-atomic ratio:

\begin{equation}
R_{\rm{mol}}(r,t)\,=\,{\Sigma_{{{\rm{H}}_2}}}(r,t)/{\Sigma_{{\rm{H{_I}}}}(r,t)}\,=\,
{\left[ {P_{\rm h}\left( r,t\right)/{P_0}} \right]^{\alpha_P} },
\label{eq:BRh2}
\end{equation}
where $P_{\rm h}(r,t)$ is the mid-plane pressure of the
ISM. $P_{0}$ and $\alpha_{\rm P}$ are constants derived from
the observations. We adopt $P_{0}/k\,=\,1.7\times10^{4}\,\rm{cm}^{-3}\,\rm{K}$ and
$\alpha_{\rm P}\,=\,0.8$ \citep{Leroy2008}.

The mid-plane pressure of the ISM in disc galaxies can be expressed as
\citep{Elmegreen1989,Leroy2008}:
\begin{equation}
P_{\rm h}\left( r,t \right)\,=\,\frac{\pi }{2}{\rm G}{\Sigma _{{\rm{gas}}}}\left( r,t \right)\left[ {{\Sigma _{{\rm{gas}}}}\left( r,t \right)
+ \frac{c_{\rm gas}}{c_{\rm *}}{\Sigma _*}\left( r,t \right)} \right],
\label{eq:elmegreenpressure}
\end{equation}
where $\rm G$ is the gravitational constant, and $c_{\rm{gas}}$ and
$c_{\rm{*}}$ are the (vertical) velocity dispersions of gas and
stars, respectively. Observations reveal that $c_{\rm{gas}}$ is a
constant along the disc and we adopt $c_{\rm{gas}}\,=\,11\rm\,km\,s^{-1}$
\citep{Ostriker2010}, while $c_{\rm{*}}$ is estimated as
$c_{\rm *}\,=\,\sqrt{\frac{2\pi {\rm G}R_{\rm d}}{7.3}}\Sigma _*(r,t)^{0.5}$,
see Appendex B.3 of \citet{Leroy2008} for more details.

The feedback from supernova gives rise to galactic outflows
when the thermal energy of a galaxy exceeds the binding energy of gas.
NGC\,7793 is a low-mass disc galaxy with shallower gravitational potential,
making it more susceptible to losing its ISM
\citep{Garnett2002, Tremonti2004, Hirschmann2016, Lian2018a, Spitoni2021b}. Thus,
the gas-outflow process is important during the evolution and SFH of NGC\,7793.
We assume that the outflowing gas does not fall again to the disc, and its
metallicity is equal to that of ISM at the time the outflow process is launched
\citep{Chang2010, Kang2012, Kang2016, Kang2017, Ho2015}. The gas outflow rate
$f_{\rm out}(r,t)$ is proportional to $\Psi(r,t)$ \citep[see ][]{Recchi2008, Spitoni2015, Spitoni2020}:
\begin{equation}
  f_{\rm out}(r,t)=b_{\rm out}\Psi(r,t).
\label{eq:outflow}
\end{equation}
where $b_{\rm out}$ is the gas outflow efficiency, and it is the other free
parameter in our model. The assumption $f_{\rm out}(r,t)$ proportional to $\Psi(r,t)$ is
reasonable if supernova and winds of massive stars are the main drivers of the
galactic wind.

In summary, the gas infall timescale $\tau$ and the gas outflow
efficiency $b_{\rm out}$ are two free parameters in our model.
The best-fitting model predicted $\tau$ and $b_{\rm out}$ obtained
in Sect.\,\ref{sec:result_r} will be presented in Table\,\ref{tab:best}.
Moreover, a degeneracy between the yield $y$ and the outflow parameter
$b_{\rm out}$ exists in that the model adopting a higher $y$ needs a
larger $b_{\rm out}$ to reproduce the observed radial metallicity distribution.
Thanks to the fact that the reasonable range of y is small for a
fixed IMF \citep{Vincenzo2016} compared with the large possible rang of
$b_{\rm out}$, we can constrain $b_{\rm out}$ by using the observed metallicity
gradient. It should be emphasize that, although
the accurate value of free parameters in our best-fitting model
may change a little, our results of the main trends of the
SFHs of NGC\,7793 are robust.

%--------------------------------------------------------------------

\section{Observations}
\label{sec:observe}

\begin{table*}
\caption{Basic properties of NGC\,7793, NGC\,2403, NGC\,300 and M\,33.}
\label{Tab:obs1}
\begin{center}
\begin{tabular}{lllll}
\hline
\hline
Property                  &     NGC\,7793         &    NGC\,2403              &   NGC\,300                 &   M33\\
\hline
Morphology                &    SA(s)d$^{a, b}$    &  SAB(s)cd$^{a, b}$    &  SA(s)d$^{a, b}$       & SA(s)cd$^{a, b}$  \\
Distance\,(Mpc)             &    $3.9^{c}$          &  3.2$^{d,e}$          &  2.0$^{d}$             & 0.8$^{f}$    \\
$M_{\rm B}\,(\rm mag)$      &    $-18.7^{g}$        &  $-18.6^{h}$          &  $-17.66^{i}$          & $-18.4^{i}$  \\
$M_{\rm K}\,(\rm mag)$      &    $-19.40^{j}$       &  $-21.3^{j} $         &  $-20.1^{j}$           & $-20.4^{j}$  \\
Scale-length\,(kpc)         &    $1.3^{k}$          &  1.6$^{k}$            &  1.29$^{l}$            &  1.4$^{l}$   \\
Rotation velocity\,($\rm km\,s^{-1}$)  & 116$^{m}$  &  136$^{m}$            &  91$^{m}$              &  110$^{m}$   \\
Stellar mass\,($10^{9}\rm M_{\odot}$)  & 3.16$^{k}$ &  5.01$^{k}$           &  1.93$^{l}$            &  4.5$^{l}$   \\
\hline
\end{tabular}\\
\end{center}
Refs: (a) NED; (b) \citet{1991S&T....82Q.621D}; (c) \citet{Karachentsev2004};
(d) \citet{Dalcanton2009}; (e) \citet{Freedman2001}; (f) \citet{Williams2009};
(g) \citet{Prugniel1998}; (h) \citet{Lee2011}; (i) \citet{Gogarten2010};
(j) \citet{Jarrett2003}; (k) \citet{Leroy2008}; (l) \citet{MM2007}; (m) \citet{Garnett2002}.
\end{table*}

A successful chemical evolution model of NGC\,7793, especially one
involving free parameters, should reproduce as many observed constraints
as possible, including both local (concerning the radial profiles) and
global (concerning the whole disc) constraints.
In that sense, the observed present-day cold gas, SFR, sSFR and metallicity
distributions provide crucial constraints on the model. Thus, we summarise
the current available observations for the disc of NGC\,7793 in this section.

\subsection{Radial distributions of gas mass, SFR, sSFR and metallicity}
\label{sec:gasstar}

The stratification of atomic hydrogen (HI) of NGC\,7793 is obtained from the Very Large
Array (VLA) maps of the National Radio Astronomy Observatory (NRAO)
\citep{Walter2008}, and the radial distribution HI mass surface density
($\Sigma_{\rm HI}$) of NGC\,7793 is taken from \citet{Leroy2008}.
The radial SFR surface density ($\Sigma_{\rm SFR}$) along the disc of
NGC\,7793 is estimated by the combination of FUV with 24\,$\mu\rm m$
maps \citep{Leroy2008} and by the FUV emission corrected by the WISE 22$\mu\rm m$
\citep{Casasola2017}.

sSFR, defined as SFR per unit of stellar mass, represents the
ratio of young to old stars and shows what fraction of total star formation
has been occurred recently. \citet{MM2007} derived the radial sSFR profiles
for NGC\,7793 from GALEX and 2MASS (FUV$-K$) colour profiles after a
proper SFR calibration of the UV luminosity and $K-$band mass-to-light ration
are adopted. \citet{Smith2021} calculated the radial sSFR profiles for NGC\,7793
by using broad-band indicators for SFR (FUV$\,+\,24\,\mu\rm m$ flux) and stellar
mass ($3.6\,\mu\rm m$ flux).

The present-day distribution of metallicity within an individual galaxy
is the outcome of a complex pattern of evolution, advancing through infall of
metal-free gas, star formation, energy feedback from stars and metal-enriched
outflows. Each of these processes, to various extents, reshuffles the metals
present in and around galaxy discs. Thus, the radial metallicity distribution
of galaxies places important constraints on the galaxy chemical model
\citep[][and references therein]{BP2000, Ho2015, Kubryk2015, Kang2012, Kang2016, Kang2017, Kang2021, Bresolin2019}.
The radial metallicity gradient has been studied for a long time, starting
with the pioneering work by \citet{Aller1942}.
Most spiral galaxies in the local Universe exhibit negative metallicity
gradients within their optical radius, i.e. the centre of a galaxy
has a higher metallicity than the outskirts
\citep[e.g. ][]{Zaritsky1994, Rupke2010, Moustakas2010, Sanchez2014,
Pilyugin2014, Kudritzki2015, Ho2015, Bresolin2019}.

Since oxygen is the most abundant metal by mass in the Universe, it is
a good tracer for the total metal content. Oxygen is also an element for
which the IRA approximation is appropriate, therefore oxygen abundance
(i.e. ${\rm 12+log(O/H)}$) will be used to represent the metallicity of
NGC\,7793 and adopt the solar value as $\rm 12+log(O/H)_{\odot}\,=\,8.69$
\citep{Asplund2009} throughout this work. The absolute gas-phase metallicity
depends on the calibrations used, and the classical T$_{\rm e}$ method is
generally considered to provide the most reliable oxygen abundances in
HII-regions \citep{Izotov2006, Pilyugin2014}. Thus, we use the oxygen
abundances in HII-regions determined by the T$_{\rm e}$  method
to constrain the model.
The radial distribution of metallicity from HII-regions emission line analysis along
the disc of NGC\,7793 have been observed and quantified by many works
\citep{McCall1985, Zaritsky1994, Moustakas2010, Bibby2010, Pilyugin2014, Stanghellini2015},
and the data from \citet{Bibby2010}, \citet{Pilyugin2014} and \citet{Stanghellini2015}
are used to constrain the model adopted in this work.
The linear radial metallicity gradient depends on the galaxy distance,
so we have scaled the metallicity gradient to the distance of $3.9\rm\,Mpc$.
The metallicity gradient of NGC\,7793 is $-0.07\pm0.019\,\rm dex\,kpc^{-1}$
and $-0.36\pm0.10\,\rm dex\,R_{25}^{-1}$ \citep{Bibby2010},
$-0.0662\pm0.0104\,\rm dex\,kpc^{-1}$ and $-0.350\pm0.055\,\rm dex\,R_{25}^{-1}$
\citep{Pilyugin2014},
$-0.054\pm0.019\,\rm dex\,kpc^{-1}$ and $-0.286\pm0.102\,\rm dex\,R_{25}^{-1}$
\citep{Stanghellini2015}.

All the aforementioned observed radial distributions of HI mass, SFR, sSFR and metallicity
will be displayed in Fig.\,\ref{fig:model_result} to constrain the model
for searching for the best-fitting model of NGC\,7793.

\begin{table}
\caption{Global observational data for the disc of NGC\,7793.}
\label{Tab:obs2}
\begin{center}
\begin{tabular}{lll}
\hline
Property      &    Value                                       &   References   \\
\hline
HI mass       & $\sim(1.05-1.26)\,\times10^{9}~\rm M_{\odot}$  & 1,2     \\
H$_2$ mass    & $\sim2.0\,\times10^{8}~\rm M_{\odot}$          & 2       \\
Gas fraction  & $\sim0.27-0.33$                                &         \\
Total SFR     & $\sim0.235-0.52\,\rm M_{\odot}\,yr^{-1}$       & 1, 3, 4, 5, 6   \\
sSFR          & $\sim(8.06-9.74)\times10^{-11}\,\rm yr^{-1}$   & 2, 7    \\
${\rm 12+log(O/H)_{R_{\rm e}}}$      & $\sim8.31-8.87$         & 3, 8, 9, 10 \\
\hline
\end{tabular}\\[1mm]
\end{center}
%$^{\rm a}$ $R_{\rm e}$ is defined as the radius at which the integrated
%flux is half of the total one, and it is equal to 1.685 times the
%radial scale-length $R_{\rm d}$ of the disc.\\
Refs: (1) \citet{Leroy2008}; (2) \citet{Muraoka2016}; (3) \citet{Bibby2010};
(4) \citet{Kennicutt2011}; (5) \citet{Skibba2011}; (6) \citet{Calzetti2015};
(7) \citet{MM2007}; (8) \citet{Moustakas2010}; (9) \citet{Pilyugin2014};
(10) \citet{Stanghellini2015}.
\end{table}

\subsection{Global properties of gas mass, SFR, sSFR and metallicity}
\label{sec:global}

The atomic hydrogen (HI) gas mass of NGC\,7793 is $M_{\rm HI}\,=\,1.26\,\times10^{9}~\rm M_{\odot}$
\citep{Leroy2008} and $M_{\rm HI}\,=\,1.05\,\times10^{9}~\rm M_{\odot}$
\citep{Muraoka2016}. The molecular hydrogen ($\rm H_{2}$)
gas mass of NGC\,7793 is $M_{\rm H2}\,=\,2.0\times10^{8}{\rm M_{\odot}}$
\citep{Muraoka2016}. As a result, the gas fraction ($f_{\rm gas}$) of NGC\,7793
can be easily calculated through
$f_{\rm gas}\,=\,\frac{M_{\rm HI}+M_{\rm H2}}{M_{\rm HI}+M_{\rm H2}+M_*}$,
and the value of $f_{\rm gas}$ is $\sim0.27-0.33$.

The current global SFR of NGC\,7793 disc has been measured by using different
tracers, $0.235\,\rm M_{\odot}\,yr^{-1}$ from a combination of FUV with
$24\,\rm \mu m$ maps by \citet{Leroy2008}, $0.45\,\rm M_{\odot}\,yr^{-1}$ from
the observed $\rm H\alpha+$[N{\sc ii}]$\rm \lambda6583$ flux by
\citet{Bibby2010}, $0.26\,\rm M_{\odot}\,yr^{-1}$ from a
combination of $\rm H\alpha$ with $24\,\rm \mu m$ maps by \citet{Kennicutt2011},
$0.363\,\rm M_{\odot}\,yr^{-1}$ from a combination of FUV with total
infrared (TIR) maps by \citet{Skibba2011}, $0.52\,\rm M_{\odot}\,yr^{-1}$
calculated from the GALEX FUV corrected for dust attenuation
\citep{Calzetti2015}, and $0.251\,\rm M_{\odot}\,yr^{-1}$ obtained from hybrids
of FUV and $22\,\rm \mu m$ maps by \citet{Leroy2019}.

The value of sSFR for NGC\,7793 is $\sim(8.36-9.74)\times10^{-11}\,\rm yr^{-1}$
derived by using FUV$-K$ colour \citep{MM2007} and $8.06\times10^{-11}\,\rm yr^{-1}$
estimated by \citet{Muraoka2016}.

The value of metallicity at the effective radius of the
disc ${\rm 12+log(O/H)_{R_{\rm e}}}$ is used as a surrogate for the
mean metallicity of a galaxy \citep{Zaritsky1994, Sanchez2013}.
$R_{\rm e}$ is equal to 1.685 times the radial scale-length $R{\rm d}$
of the disc \citep{Vaucouleurs1978} and is basically corresponding to
$0.4R_{25}$ \citep{Sanchez2013} in the local Universe, i.e.
$R_{\rm e}\,=\,1.685R_{\rm d}\,=\,0.4R_{\rm 25}$.
The value of mean metallicity for the disc of NGC\,7793 is $8.466\pm0.05$
\citep{Bibby2010}, $8.31\pm0.02$ \citep[calibration of][]{KK04} and
$8.87\pm0.01$ \citep[calibration of][]{PT05} \citep{Moustakas2010},
$8.378\pm0.02$ \citep{Pilyugin2014} and $8.445\pm0.063$
\citep{Stanghellini2015}, respectively.

The above-mentioned global observed properties for the disc of NGC\,7793 are
summarised in Table \ref{Tab:obs2}.

\section{Results and discussion}
\label{sec:result}

%\subsection{Modeling the gas-phase and stellar metallicity}
%\label{sec:model_Z}

\begin{figure*}
  \centering
  \includegraphics[angle=0,scale=0.8]{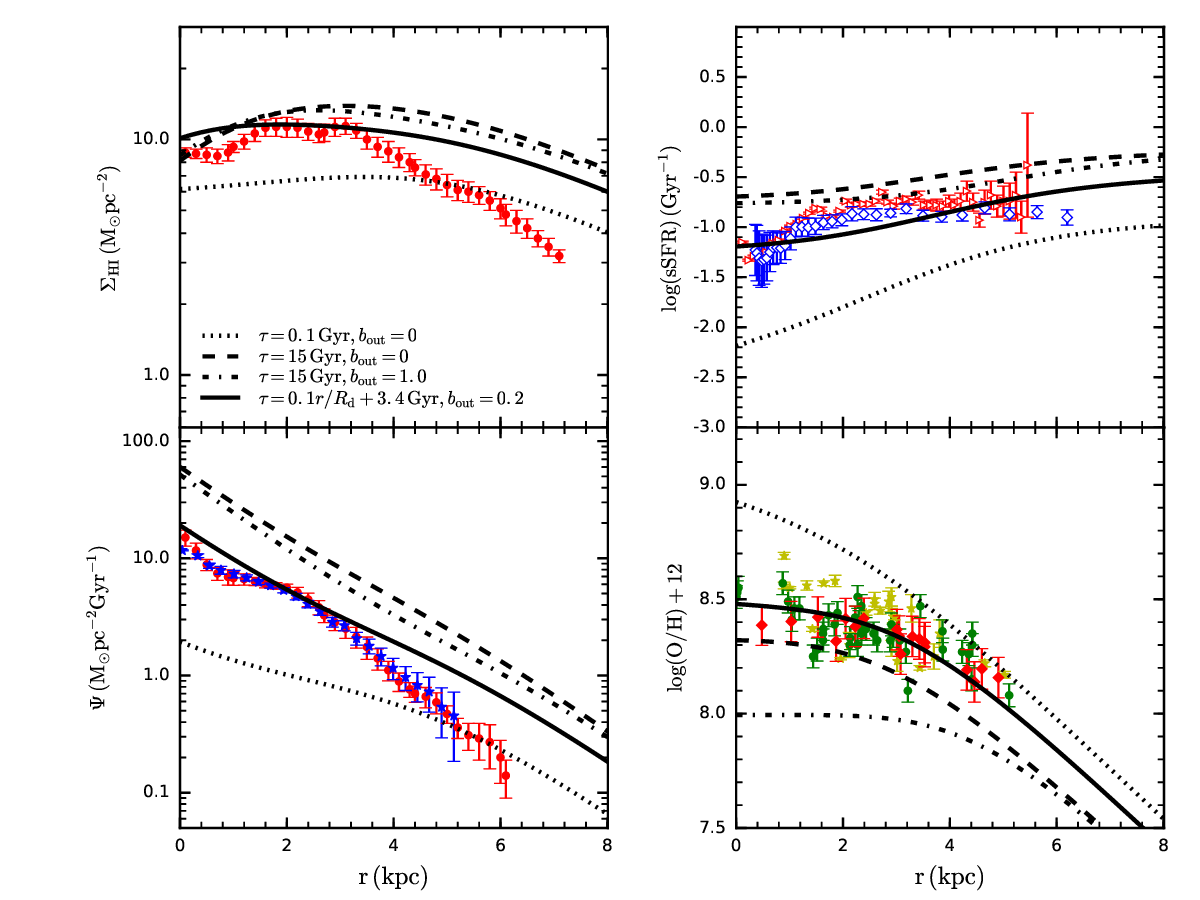}
    \caption{Comparisons of the model predictions with the observations.
   Different line types correspond to different parameter groups:
   dotted lines $(\tau, b_{\rm out})=(0.1\,{\rm Gyr}, 0)$, dashed
   lines $(\tau, b_{\rm out})=(15\,{\rm Gyr}, 0)$, dot-dashed lines
   $(\tau, b_{\rm out})=(15\,{\rm Gyr}, 1.0)$, solid lines
   $(\tau, b_{\rm out})=(0.1r/{R_{\rm d}}+3.4\,{\rm Gyr}, 0.2)$.
   On the left-hand side, the radial profiles of H{\sc i} mass and SFR
   surface density are shown in the top and bottom panels, respectively;
   On the right-hand side, the radial profiles of sSFR and metallicity are
   displayed in the top and bottom panels, respectively.
   Different symbols are corresponding to the
   observed data taken from different works.
   H{\sc i} data from \citet{Leroy2008} are shown by red filled circles.
   SFR data obtained from \citet{Leroy2008} are denoted as red filled circles,
   while those from \citet{Casasola2017} are displayed by blue filled asterisks.
   sSFR data from \citet{MM2007} and \citet{Smith2021} are denoted as
   red open triangles and red open diamonds, respectively.
   ${\rm 12+log(O/H)}$ data taken from \citet{Bibby2010}, \citet{Pilyugin2014}
   and \citet{Stanghellini2015} are separately plotted by yellow filled asterisks,
   green filled circles and red filled diamonds.
}
\label{fig:model_result}
\end{figure*}

In this section, we firstly investigate the influence of the free parameters on model
predictions and search for the best-fitting model for NGC\,7793.
Secondly, we study the properties of stellar populations along the disc of NGC\,7793.
Finally, to search for clues to the processes that drive the disc formation
scenarios and stellar mass growth histories, we compare the best-fitting model predicted
growth history of NGC\,7793 with those of M33, NGC\,300 and NGC\,2403, which have been
separately studied in our previous work \citep{Kang2012, Kang2016, Kang2017}.

\subsection{Radial distributions}
\label{sec:result_r}

The aforementioned free parameters in the model are the gas infall timescale
$\tau$ and the outflow efficiency $b_{\rm out}$. We explore the influence of these
two free parameters on the model predictions through comparing the model predictions
with observations. The dotted, dashed and dot-dashed lines in Fig.\,\ref{fig:model_result}
separately denote the predictions of the model adopting $(\tau, b_{\rm out})=(0.1\,{\rm Gyr}, 0)$,
$(\tau, b_{\rm out})=(15\,{\rm Gyr}, 0)$ and $(\tau, b_{\rm out})=(15\,{\rm Gyr}, 1.0)$.
The left-hand side of Fig.\,\ref{fig:model_result} shows the radial profiles of
H{\sc i} mass (top) and SFR (bottom) surface density, and the right-hand side of
Fig.\,\ref{fig:model_result} displays the radial profiles of sSFR (top) and
metallicity (bottom). The details of the observed data were presented in
Sect.\,\ref{sec:observe}.

The comparison of the three models in Fig.\,\ref{fig:model_result}
shows that the outflow efficiency has a strong influence on
metallicity but only a small effect on HI column density, SFR and
sSFR, while the gas infall timescale is important for the radial
distribution of all four quantities. This is mainly due to the fact that, during the
whole evolutionary history of a galaxy, the infalling gas supplies the reservoir
for star formation, while the outflowing gas takes a fraction of metals away
from the disc. Figure\,\ref{fig:model_result} also shows that the area between
the dotted line $(\tau, b_{\rm out})=(0.1\,{\rm Gyr}, 0)$ and the
dot-dashed line $(\tau, b_{\rm out})=(15\,{\rm Gyr}, 1.0)$ brackets almost the whole region of
the observed data, which implies that it is possible to construct a model that
can reproduce the main observed features of NGC\,7793 disc.

The right bottom panel of Fig.\,\ref{fig:model_result} exhibits negative
metallicity gradient, i.e. the central region has a higher metallicity than
the outskirts, consistent with the disc inside-out formation scenario.
Moreover, the colour--magnitude diagrams (CMD)-derived SFH also
certifies an inside-out growth for the disc of NGC\,7793 \citep{Sacchi2019}.
Thus, following the method adopted in our previous work
\citep{Kang2012, Kang2016, Kang2017}, the form of infall timescale
is assumed to be $\tau(r)\,=\,a\times r/{R_{\rm d}}+b$,
namely that gas takes longer timescale to settle onto
the disc in the outer regions, where $a$ and $b$ are
the coefficients for $\tau(r)$. An inside-out growth
mechanism originally introduced in the theory of galaxy evolution
on the basis of chemical evolution models \citep{Larson1976, Matteucci1989} and
semianalytic model of disc galaxies in the context of dark matter cosmologies
\citep{Kauffmann1996, vandenBosch1998}. Adding the aforementioned free parameter
$b_{\rm out}$, there are now three free parameters ($a$, $b$ and $b_{\rm out}$)
in our model that should be determined.

In order to search for the best combination of free parameters ($a$, $b$ and
$b_{\rm out}$) able to reproduce the main observed features of NGC\,2403,
we use the classical $\chi^{2}$ technique by comparing the model results with
the corresponding observational data, such as the radial profiles of H{\sc i}
gas mass surface density, SFR, sSFR and $\rm 12+log(O/H)$. The
boundary conditions of $a$, $b$ and $b_{\rm out}$ for NGC\,7793
are separately assumed to be $0<\,a\,\leq3.0$, $1.0\leq\,b\leq\,5.0$
and $0.1\leq\,b_{\rm out}\leq\,0.9$.
In practise, for each pair of $a$ and $b$, we vary the value of $b_{\rm out}$ to
calculate the minimum value of $\chi^{2}$, which corresponds to the best
combination of $a$, $b$ and $b_{\rm out}$, and we obtain
($a$, $b$, $b_{\rm out}$)\,=\,(0.1, 3.4, 0.2), i.e.
$(\tau, b_{\rm out})\,=\,(0.1r/{R_{\rm d}}+3.4\,{\rm Gyr}, 0.2)$
defined as the best-fitting model, and the best combination of free parameters
are displayed in Table\,\ref{tab:best}. The best-fitting model predicted results
are plotted by solid lines in Fig.\,\ref{fig:model_result}. It can be found that
there is remarkable agreement between the model results and the NGC\,7793 observations.

\begin{figure}
   \includegraphics[angle=0,scale=0.55]{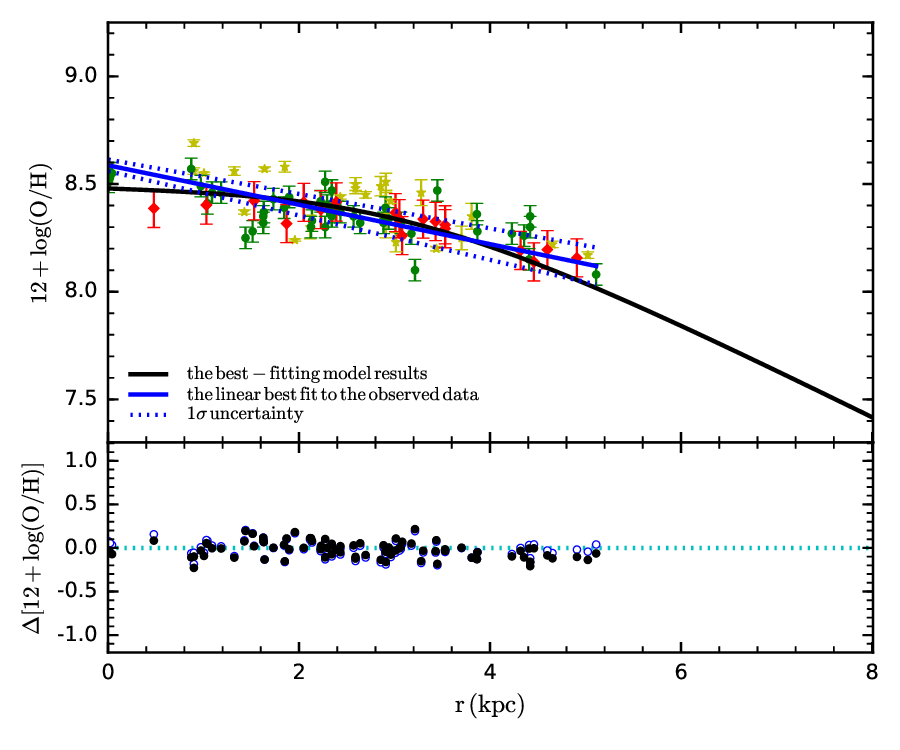}
   \caption{Radial distribution of ${\rm 12+log(O/H)}$.
   Upper panel: comparisons of the present-day radial distribution of
   $\rm 12+log(O/H)$ predicted by our best-fitting model (black solid line) with
   the observational data. Different symbols denote the
   observed metallicity data from different works: yellow filled asterisks
   \citep{Bibby2010}, green filled circles \citep{Pilyugin2014} and red filled
   diamonds \citep{Stanghellini2015}.
   The linear best fit to the observed data and $1\sigma$
   uncertainty are plotted as the blue solid and dotted lines, respectively.
   Lower panel: the black filled circles show the deviations of the observations
   from our best-fitting model predictions, derived by the black solid line minus the observed
   data in the above panel, while the blue open circles show the ones obtained from the
   linear best fit minus the observational data in the above panel.
   }
\label{Fig:Z_dev}
\end{figure}

The right lower panel of Fig.\,\ref{fig:model_result} shows that
the best-fitting model predicted radial metallicity
distribution does not show a constant gradient of $\rm 12+log(O/H)$. Instead there
is a bend in the radial metallicity distribution. This is mainly due to the fact that we use
the observational data, such as the radial distributions of gas mass surface
density, SFR surface density, sSFR and 12+log(O/H), to constrain the model and to
search for the best-fitting model which can simultaneously reproduce these observational
data for NGC7793.
In order to further state the best-fitting model can nicely describe the formation and
evolution of NGC\,7793, comparisons of the best-fitting model predicted present-day
$\rm 12+log(O/H)$ distributions (black solid line) with the observed data (points)
are displayed in the upper panel of Fig.\,\ref{Fig:Z_dev}. The linear best fit and
the $1\sigma$ uncertainty of the observed data are also shown in this panel by the
blue solid line and blue dotted lines, respectively. The deviations of the observed
$\rm 12+log(O/H)$ points from our best-fitting
model predicted $\rm 12+log(O/H)$, calculated by the best-fitting model predictions minus the
observed data, are shown as the black filled circles in the lower panel of
Fig.\,\ref{Fig:Z_dev}, and the deviations of the observed metallicity date from the
corresponding linear best fit are also plotted in this panel as the blue open circles.
It can be found from the lower panel of Fig.\,\ref{Fig:Z_dev} that there is little difference
between the black solid circles and the blue open circles. The mean deviation values of the
black filled circles and the blue open circles are 0.019 and 0.015, respectively.
Figure\,\ref{Fig:Z_dev} reinforces our results that the best-fitting model includes
and describes reasonably the crucial ingredients of the main physical processes
that regulate the formation and evolution history of NGC\,7793.

\subsection{Stellar populations along the disc}

\begin{figure}
  \centering
  \includegraphics[angle=0,scale=0.55]{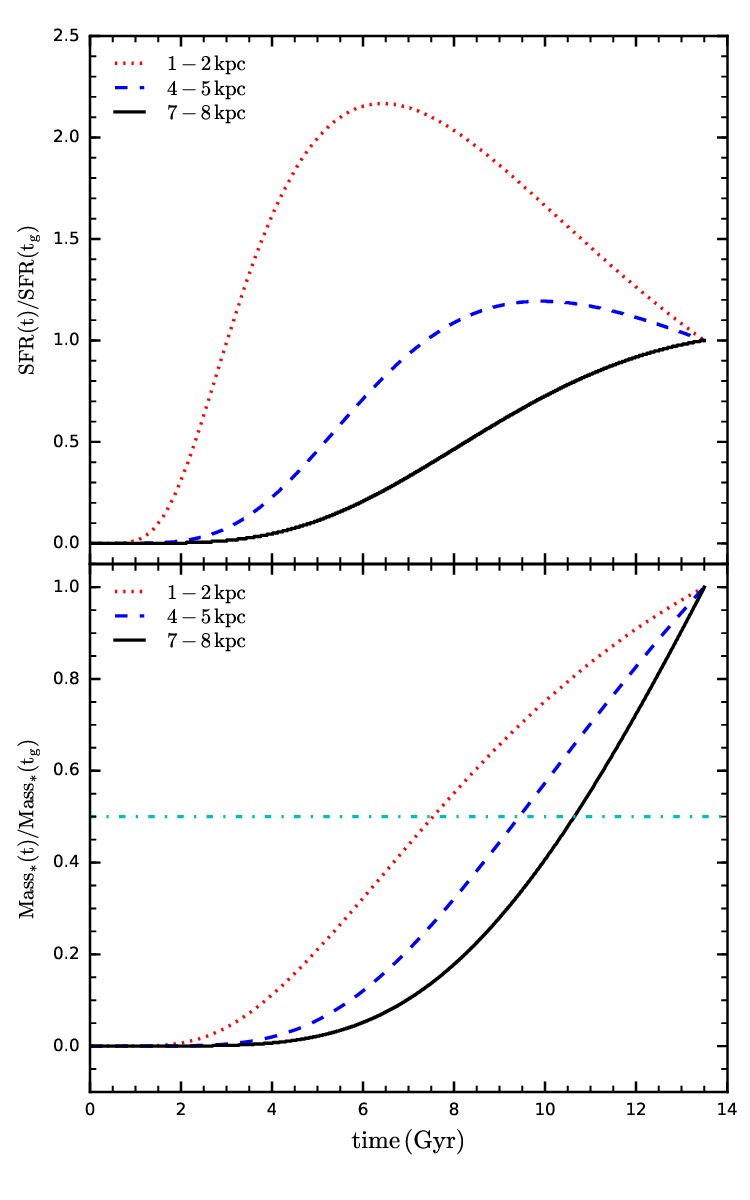}
    \caption{Star formation histories (upper panel) and relative stellar mass
    growth (lower panel) of three regions at different galactocentric distances
    (1-2\,kpc, 4-5\,kpc, 7-8\,kpc) for NGC\,7793 disc.
    Both the SFRs and stellar masses are
    normalized by their present-day values. The horizontal
    dash-dotted line in the lower panel remarks when each component
    achieves 50\% of its final value.
}
\label{fig:formation_scenario}
\end{figure}

To investigate the stellar populations along the disc of NGC\,7793,
Figure\,\ref{fig:formation_scenario} demonstrates the best-fitting
model predicted SFHs (upper panel) and stellar mass growth curves (lower panel) of
three regions at different galactocentric distances for NGC\,7793.
Different line types represent different regions along the disc, red dotted line for
$1-2\,\rm kpc$, blue dashed line for $4-5\,\rm kpc$ and black solid line
for $7-8\,\rm kpc$. The horizontal cyan dash-dotted line in the lower panel
marks when each component reaches 50\% of its final value. Both the SFRs and stellar masses
in Fig.\,\ref{fig:formation_scenario} are normalized by their present-day values.
The upper panel of Fig.\,\ref{fig:formation_scenario} shows that
the peak of the SFH moves to later time from internal to more external
regions of the disc, and regions at larger radii have more extended SFHs.
The lower panel demonstrates that the stellar mass growth rate in the
inner region is faster than that in the outer region.
Figure\,\ref{fig:formation_scenario} shows clear signatures of inside-out
growth for NGC\,7793 disc.

\begin{figure}
  \centering
  \includegraphics[angle=0,scale=0.55]{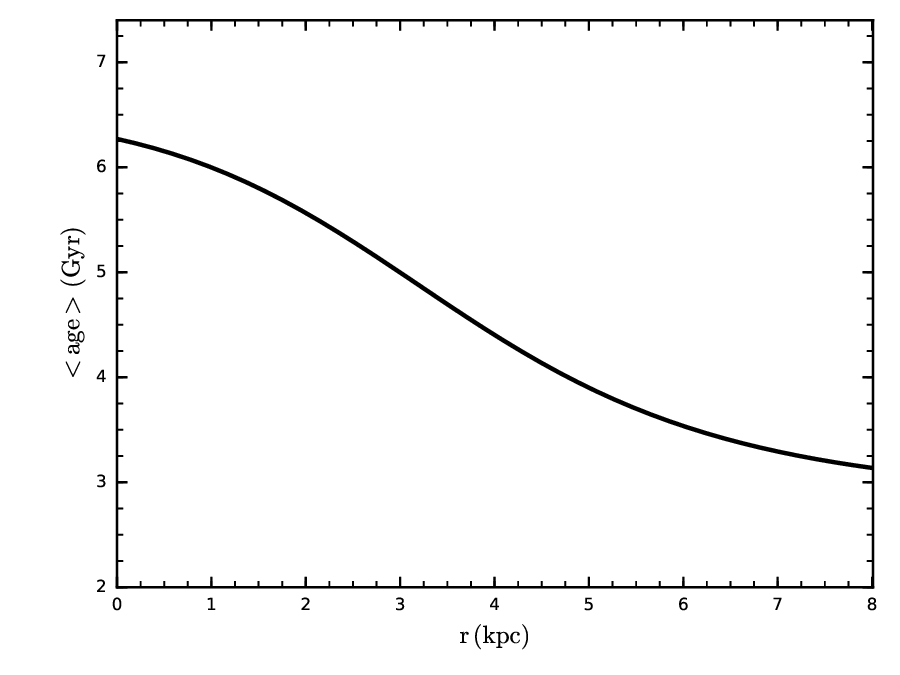}
    \caption{Radial profile of mean stellar age along the disc of NGC\,7793
    predicted by its own best-fitting model.
}
\label{fig:age}
\end{figure}

If the disc of NGC\,7793 does form "inside-out", one would expect to see
negative radial gradient in age. Figure\,\ref{fig:age} shows the mean
stellar age along the disc of NGC\,7793 predicted by the best-fitting model
with solid line. It can be found that the stellar population
in the inner region of the disc is older than that in the outer region, that is, the
inner parts of the disc have a higher percent of old stars than in the outer parts.
The mean stellar age which decreases with radius is indeed consistent with the
inside-out growth scenario for the disc formation. In this picture, the inner
regions of NGC\,7793 are built up at earlier times than outer parts, and as a result
contain on average older stars than outermost regions.
The results presented in both Fig.\,\ref{fig:formation_scenario} and Fig.\,\ref{fig:age}
are indicative of an inside-out formation scenario for the disc of NGC\,7793, in line
with the observed results derived by \citet{MM2007} and \citet{Sacchi2019}.

\begin{figure}
  \centering
  \includegraphics[angle=0,scale=0.55]{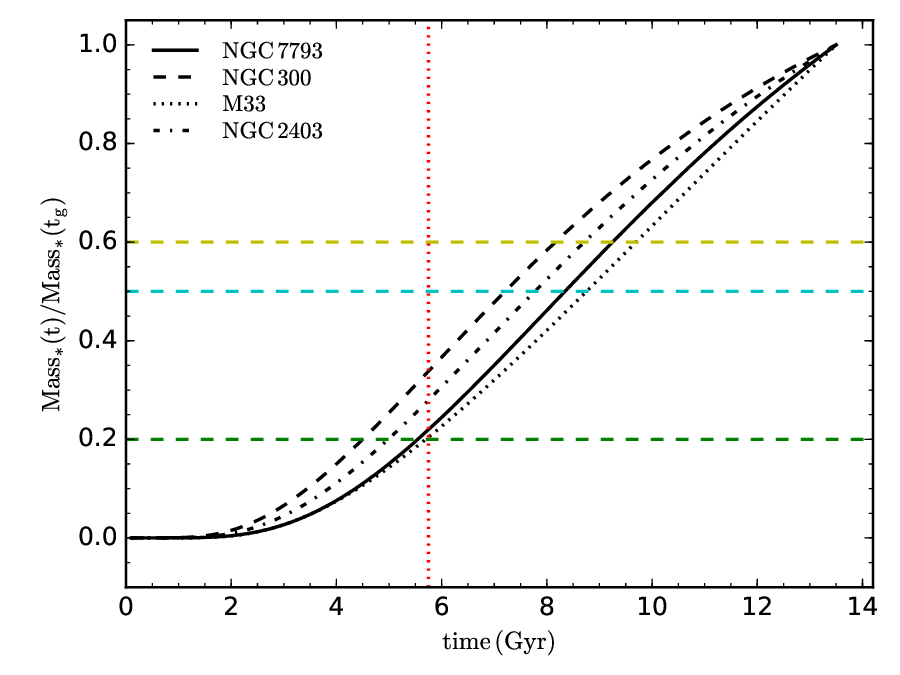}
    \caption{Stellar mass growth histories of NGC\,7793 (solid line),
    NGC\,300 (dashed line), M33 (dotted line) and NGC\,2403 (dash-dotted line)
    predicted by their own best-fitting models. Stellar masses
    are normalized to their present-day values. The horizontal dashed
    lines mark when the stellar mass reaches 20\%(green), 50\%(cyan) and
    60\%(yellow) of its final value, while the vertical red dotted
    line denotes the galaxy evolutionary age at
    $t\,=\,5.75\,\rm Gyr\,({\rm i.e. z\,=\,1}\,)$.
    }
  \label{Fig:stellar}
\end{figure}

The solid line in Fig.\,\ref{Fig:stellar} displays the best-fitting model
predicted growth history of stellar mass for NGC\,7793, and stellar
mass is normalized to its present-day value. To make the stellar mass
growth rate more visible, the horizontal dashed lines separately mark
when the stellar mass reaches 20\% (green), 50\% (cyan) and 60\%(yellow)
of its final value. The vertical red dotted line denotes the galaxy
evolutionary age at $t\,=\,5.75\,\rm Gyr\,({\rm i.e. z\,=\,1}\,)$.
It can be found that the resulting stellar mass growth
history displays a relatively smooth build-up of its stellar mass. The
best-fitting model results also show that about 80\% of the stellar mass
of NGC\,7793 is assembled within the last 8\,Gyr and 40\% within the last
4\,Gyr, consistent with what found by the CMD-derived SFH for NGC\,7793 in
\citet{Radburn-Smith2012} and \citet{Sacchi2019}. Other line types in this
Figure will be discussed in the Sect.\,\ref{subsec:stellarSFR_t}.

\subsection{Comparison with other three disc galaxies}

\begin{table*}
\caption[best]{The main input properties and parameters of the best-fitting
models for NGC\,7793, NGC\,2403, NGC\,300 and M33.}
\begin{center}
\begin{tabular}{lllllll}
\hline
Individual                                  &  & &  NGC\,7793   &   NGC\,2403     &      NGC\,300     &     M33        \\
\hline
Input physical properties &  &  $M_{\rm \ast}$($10^{9}\rm M_{\odot}$)  &   3.16  &5.0    &    1.928   & 4.0   \\
 &  &  $R_{\rm d}$ (kpc)              &  1.3   &   1.6    &        1.29       & 1.4                   \\
Star formation law         &     $\Psi(r,t)\,=\,\Sigma_{\rm H_{2}}(r,t)/t_{\rm dep}$ &   $t_{\rm dep}\,(\rm Gyr)$                     &  1.9   &   1.9     &        1.9        &   0.46     \\
Infall rate  &    $f_{\rm{in}}(r,t)\,\propto\,t\cdot e^{-t/\tau}$    &    $\tau(r)$\,(Gyr)     & $0.1r/{R_{\rm d}}+3.4$   &   $0.2r/{R_{\rm d}}+3.2$    &$0.35r/{R_{\rm d}}+2.47$   &  $r/{R_{\rm d}}+5.0$  \\
Outflow rate    &   $f_{\rm out}(r,t)\,=\,b_{\rm out}\Psi(r,t)$         &   $b_{\rm out}$            &   0.2   &    0.6          &         0.9        &  $0.5$         \\
\hline
\end{tabular}
\end{center}
%Note:$^{\rm a}$ The solar oxygen abundance value is from \citet{asplund09}.
\label{tab:best}
\end{table*}

NGC\,7793, M33, NGC\,300 and NGC\,2403 are four bulgeless disc galaxies
with similar morphology and stellar mass, and the chemical evolution and
SFHs of NGC\,300, M33 and NGC\,2403 have already been studied in our
previous works \citep[see][]{Kang2012, Kang2016, Kang2017}. In this section,
we will compare the best-fitting model predicted SFH of NGC\,7793
with those of M33, NGC\,300 and NGC\,2403 to search for clues to the processes
that drive the disc formation scenarios and stellar mass growth histories.
Table\,\ref{tab:best} summarises the input properties (including the
total stellar mass $M_{\rm \ast}$ and the scale-length $R_{\rm d}$), and the
best-fitting parameters for the main ingredients of the models, such as
the star formation law, the gas infall rate and outflow rate. Physical details of
the model description and searching for the best-fitting models
for M33, NGC\,300 and NGC\,2403 are separately in
\citet{Kang2012, Kang2016, Kang2017}.

\subsubsection{Evolution of stellar mass and SFR}
\label{subsec:stellarSFR_t}

\begin{figure}
  \centering
  \includegraphics[angle=0,scale=0.55]{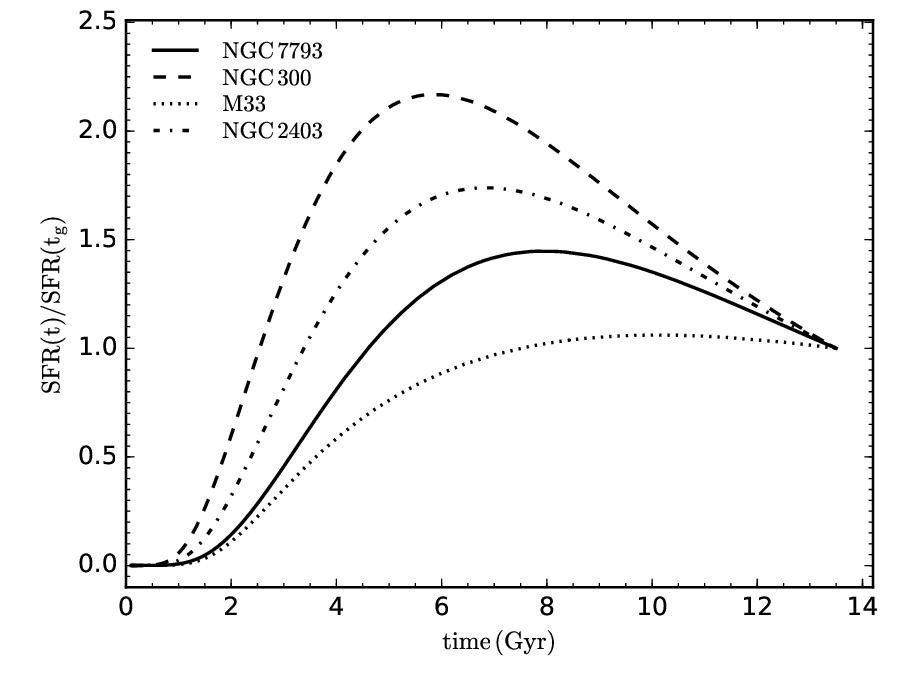}
    \caption{Evolution of SFR for NGC\,7793 (solid line),
    NGC\,300 (dashed line), M33 (dotted line) and NGC\,2403 (dash-dotted line)
    predicted by their own best-fitting models. SFRs are normalized to
    their present-day values.
    }
  \label{Fig:SFR_evolution}
\end{figure}

Figure\,\ref{Fig:stellar} plots the best-fitting model predicted
growth history of stellar mass for NGC\,7793 (solid line), NGC\,300
(dashed line), M33 (dotted line) and NGC\,2403 (dash-dotted line)
and stellar masses is normalized to their present-day values.
It can be found from Fig.\,\ref{Fig:stellar} that they have also been steadily
assembling their stellar mass. Our results are in qualitative agreement with
the results in Fig\,17 (bottom) of \citet{Sextl2023}, who used the  galaxy evolution
model in \citet{Kudritzki2021a} and found the galaxy stellar masses
at $10\,\rm Gyr$ ago are significantly smaller than the final stellar masses.
Figure\,\ref{Fig:stellar} also shows that about
78 percent (NGC\,7793), 66 percent (NGC\,300), 72 percent (NGC\,2403) and 79 percent
(M33) galaxy's stars were formed between $z\,=\,1$ and $z\,=\,0$, that is, these
four galaxies accumulated more than 50 percent of their total stellar
mass within the past $\sim\rm 8\,Gyr$ (i.e. $z\,=\,1$), supported
by the previous results that late-type galaxies with stellar mass
$M_{*}\,<\,10^{11}\,{\rm M}_{\odot}$ appear to assembled most of their
stellar mass at $z\,<\,1$ \citep{Leitner2011, Sachdeva2015},
since stellar mass contributes to the majority of the total
baryonic budget at most epochs ($z\,=\,1$).

The best-fitting model predicted SFHs of these four galaxies are
plotted in Fig. \ref{Fig:SFR_evolution}, and different line types are
corresponding to different galaxies: solid line for NGC\,7793, dashed line
for NGC\,300, dotted line for M33 and dash-dotted line for NGC\,2403.
Fig. \ref{Fig:SFR_evolution} shows that the SFHs of NGC\,7793, NGC\,2403 and
NGC\,300 reach their peaks around $5.5\,\sim\,8\rm\,Gyr$ ago and then slowly
drops down to its present-day value, while SFH of M33 reaches a maximum about
$4\rm\,Gyr$ ago and then stays constant until recently before declining somewhat.
The more extended SFHs of M33 further indicates that the principal time
of star formation on the discs of NGC\,7793, NGC\,2403 and NGC\,300 is earlier
than that on the disc of M33. This may be due to an HI bridge between M33 and
M31, in accordance with the statistical result of \citet{Guglielmo2015} that
galaxies of a given mass have different SFHs depending on their environment.

\subsubsection{Evolution of scale-length $R_{\rm d}$}

\begin{figure}
  \centering
  \includegraphics[angle=0,scale=0.55]{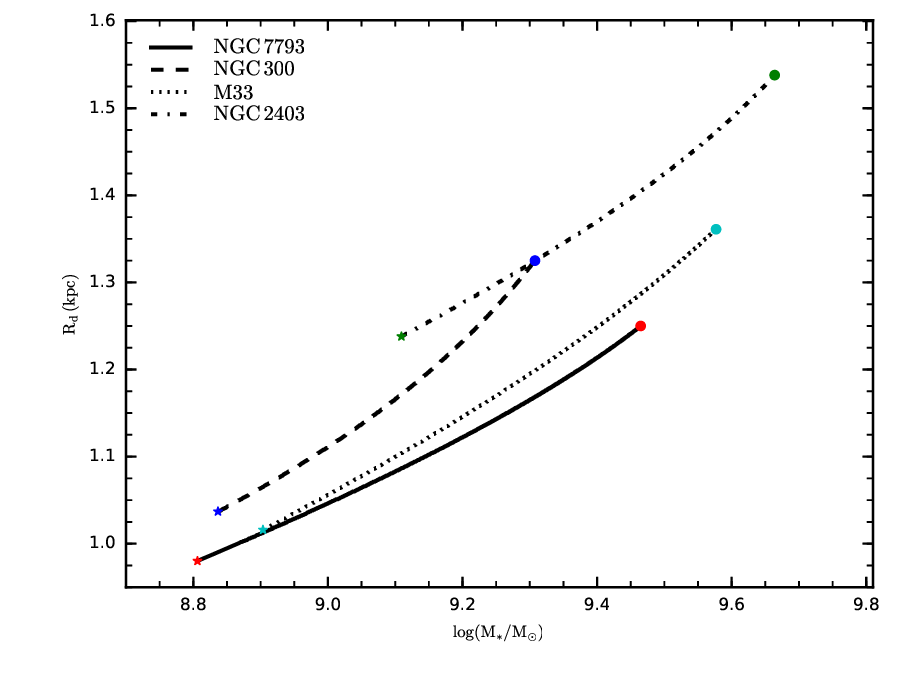}
    \caption{Evolution of the scale-length and total stellar mass
    predicted by our best-fitting models for NGC\,7793 (solid line),
    NGC\,300 (dashed line), M33 (dotted line) and NGC\,2403 (dash-dotted line).
    Each track follows the evolution of a galaxy from $z\,=\,1$ (the
    $z\,=\,1$ step is marked with asterisk) to $z\,=\,0$ (the $z\,=\,0$
    step is marked with filled circle).
}
\label{fig:M_S}
\end{figure}

Figure\,\ref{fig:M_S} plots the evolution of the scale-length and
total stellar mass (mass-size relation)
from $z\,=\,1$ to $z\,=\,0$ for NGC\,7793 (solid line), NGC\,300 (dashed line)
and M33 (dotted line), NGC\,2403 (dot-dashed line). The $z\,=\,0$ step and
$z\,=\,1$ step is marked with filled circle and asterisk, respectively.
As in \citet{MM2011}, the disc scale-length $R_{\rm d}$ at evolution time
$t$ is computed by fitting an exponential law to the total stellar mass surface
density profile, $\Sigma_*(r,t)=\Sigma_*(0,t){\rm exp}(-r/R_{\rm d}(t))$.
The model predicts $R_{\rm d}\,=\,1.25\rm\,kpc$ for NGC\,7793, which is
in agreement with its observed value $R_{\rm d}\,=\,1.30\rm\,kpc$
\citep{Leroy2008}, considering the observed uncertainties.
It can be seen from Fig.\,\ref{fig:M_S} that stellar mass and scale-length simultaneously
increase as time goes by, that is, these four spirals grow in size while
they grow in mass, in perfect agreement with the previous mass--size trend
\citep[e.g. ][]{MM2011, Pezzulli2015, Sachdeva2016}.
However, the details about how stellar discs form and grow in mass and size
are not known from first principles and significant observational effort is
still required to shed light on the missing links from structure formation to
galaxy formation.

Investigating the rate at which stellar discs grow does
shed light on the interplay between the physical processes, such as
metal-free gas infall, star formation, enriched gas outflows, and the global
evolution of disc galaxies \citep{Frankel2019}.
The specific stellar mass growth rate ($\nu_{\rm M}$) and specific radial scale-length growth rate
($\nu_{\rm R}$) represent the stellar mass and radial scale-length growth rates, defined as
$dM_{\rm \ast}/dt\,=\,\frac{M_{\rm \ast}(z=0)-M_{\rm \ast}(z=1)}{t(z=0)-t(z=1)}$
and $dR_{\rm d}/dt\,=\,\frac{R_{\rm d}(z=0)-R_{\rm d}(z=1)}{t(z=0)-t(z=1)}$,
normalized to the actual present-day value of stellar mass $M_{\rm \ast}(z=0)$
and scale-length $R_{\rm d}(z=0)$. Therefore, $\nu_{\rm R}$ between
$z\,=\,1,\,({\rm i.e.}\,t\,=5.75\,\rm Gyr)$ and $z\,=\,0\,({\rm i.e.}\,t\,=13.5\,\rm Gyr)$
calculated by their own best-fitting models are separately
$0.0278\,\rm Gyr^{-1}$ for NGC\,7793, $0.0281\,\rm Gyr^{-1}$ for NGC\,300,
$0.0252\,\rm Gyr^{-1}$ for NGC\,2403 and $0.0327\,\rm Gyr^{-1}$ for M33.
We notice that all these four disc galaxies have undergone a radial scale-length
growth of $\sim\,20\%-25\%$ since $z\,=\,1$ until now, in perfect agreement
with what found by \citet{MM2011} and \citet{Pezzulli2015}.
At the same time, $\nu_{\rm M}$ from $z\,=\,1$ to $z\,=\,0$ are
$0.1007\,\rm Gyr^{-1}$ for NGC\,7793, $0.0854\,\rm Gyr^{-1}$ for NGC\,300,
$0.0931\,\rm Gyr^{-1}$ for NGC\,2403 and $0.1016\,\rm Gyr^{-1}$ for M33.
Comparing $\nu_{\rm R}$ and $\nu_{\rm M}$, we conclude that these four disc
galaxies grow in size at $\sim\,0.30$ times the rate at which they grow in mass,
not far from the value $\sim\,0.32$ derived by \citet{Courteau2007} and
$\sim\,0.35$ found by \citet{Pezzulli2015}. Furthermore, the values of both
$\nu_{\rm M}$ and $\nu_{\rm R}$ for M33 are larger than that for other three galaxies,
indicating that recently there are more star formation occurred along the M33 disc than that
in other three galaxies. This is mainly due to the fact that H{\sc i} bridge between
M33 and M31 may be responsible for supplying cold gas for star formation in the
disc of M33 \citep{Wolfe2013}.

%\textbf{ some studies finding a significant evolution in normalization for
%$z\leq1$ (e.g. Mowla et al. 2019, ApJ, 880, 57; Nedkova et al.
%2021, MNRAS, 506, 928)}

\subsubsection{Stellar mass-Metallicity relation}

\begin{figure}
  \centering
  \includegraphics[angle=0,scale=0.55]{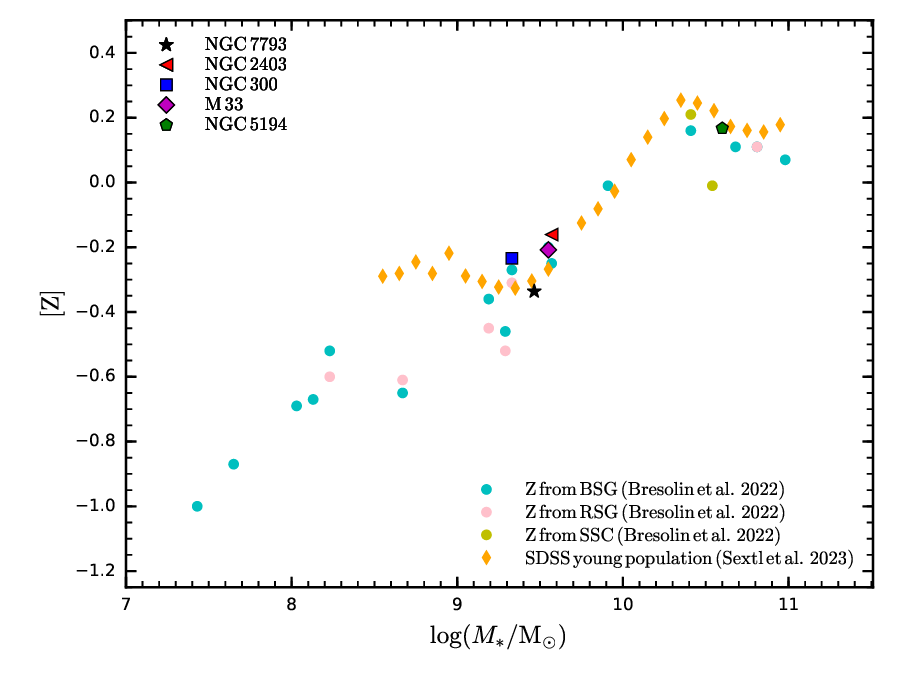}
    \caption{Comparison of the stellar mass-metallicity relation predicted by our
    best-fitting models (black star for NGC\,7793, red triangle for NGC\,2403,
    blue square for NGC\,300, magenta diamond for M33 and green pentagon for
    NGC\,5194) with that published in the literature. Solid circles are
    stellar metallicity of nearby galaxies measured from spectroscopy of individual blue
    supergiants (cyan), red supergiants (pink) and super star clusters (yellow)
    \citep[see][reference therein]{Bresolin2022}. The orange thin-diamonds are the
    MZR derived from young stellar population of SDSS star-forming galaxies
    \citep{Sextl2023}.
}
\label{fig:Mstar_Z}
\end{figure}

The shape of MZR provides
important constraints for understanding the chemical evolution of galaxies.
Figure\,\ref{fig:Mstar_Z} shows a comparison of MZR predicted by our
best-fitting models with results measured by other authors.
In order to compare our best-fitting model predicted MZR with that of young
stellar population and individual supergiant stars, we first scale the mean
metallicities of these galaxies to solar ratios, i.e.
$\rm [Z]\,=\,12+log(O/H)_{gal}\,-\,(12+log(O/H)_{\odot})\,=\,12+log(O/H)_{gal}\,-\,8.69$.
Thus, the best-fitting model predicted mean metallicities (defined as the
metallicity at the effective radius $R_{\rm e}$) are separately  $-0.336$ (NGC\,7793),
$-0.161$ (NGC\,2403), $-0.234$ (NGC\,300) and $-0.208$ (M\,33), and their values are
shown in Fig.\,\ref{fig:Mstar_Z} and denoted as black star (NGC\,7793), red triangle
(NGC\,2403), blue square (NGC\,300) and magenta diamond (M\,33).
In Fig.\,\ref{fig:Mstar_Z} we plot the MZR based on abundance data
obtained from the analysis of quantitative spectroscopy of individual objects
in $17$ nearby galaxies \citep[see Fig.\,9 and Table\,4 in][]{Bresolin2022}:
blue supergiant stars (BSG, cyan circles), red supergiant stars (RSG, pink circles)
and super star clusters (SSC, yellow circles). In this Figure, we also display the
MZR derived by \citet[][orange thin-diamonds]{Sextl2023} from the young stellar
population of SDSS local star-forming galaxies.
For comparison, the mean metallicity of NGC\,5194 \citep[][green pentagon]{Kang2015}
is also plotted in Fig.\,\ref{fig:Mstar_Z}.

It can be seen from Fig.\,\ref{fig:Mstar_Z} that our model predicted
MZR of these galaxies (including NGC\,5194) are in accordance with the MZR derived by
both young stellar population of SDSS star-forming galaxies and individual young
massive supergiant stars (including super star clusters) data.
Although our best-fitting models are constrained by the metallicities from
HII-regions emission line analysis, the model predicted MZR is compatible with
the empirical MZR measured from individual supergiant stars in nearby galaxies
and young stellar population in SDSS star-forming galaxies.

%\textbf{Some new observational metallicity data obtained from
%stellar spectral analysis of young massive blue supergiant stars
%(e.g. M33 by \citet{Liu2022}, NGC2403 by \citet{Bresolin2022}) were
%recently appeared, it would be interesting to consider these metallicities
%to constrain the chemical evolution model for these galaxies.
%It should be emphasized that, although the accurate values of
%free parameters in our best-fitting model are not unique, the main
%results may not impact our conclusion significantly. }

\subsubsection{Metallicity gradient-Stellar mass}

\begin{figure}
  \centering
  \includegraphics[angle=0,scale=0.55]{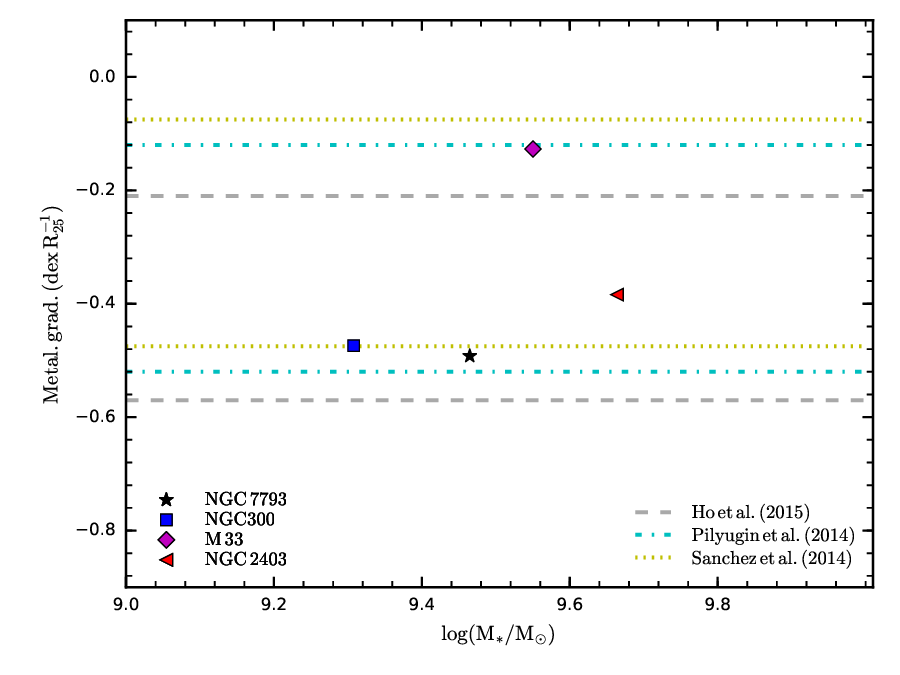}
    \caption{Metallicity gradient versus stellar mass predicted by their own
    best-fitting models when the metallicity gradients are measured in
    $\rm dex\,R_{25}^{-1}$: NGC\,7793 (black star), NGC\,2403 (red triangle),
    NGC\,300 (blue square) and M33 (magenta diamond). Different line types
    represent observed metallicity gradients from different authors: dashed
    lines \citep{Ho2015}, dotted lines \citep{Sanchez2014}, and dash-dotted lines
    \citep{Pilyugin2014}.
}
\label{fig:M_gradient}
\end{figure}

The dependence of metallicity gradient on galaxy stellar mass is a matter of
recent controversy. Isolated spiral galaxies share a characteristic metallicity gradient
%independent of morphology, the existence of a bar, absolute magnitude or mass,
when normalized to an appropriate scale-length, such as $R_{\rm d}$, $R_{\rm e}$ or
$R_{\rm 25}$ \citep{Zaritsky1994, Garnett1997,
Sanchez2014, Bresolin2015, Ho2015, Lian2018b, Pilyugin2019, Bresolin2019}.
On the other hand, \citet[][based on MaNGA]{Belfiore2017} and
\citet[][based on SAMI]{Poetrodjojo2018}
find that the metallicity gradient in terms of ${\rm dex}\,R_{\rm e}^{-1}$ steepens
with galaxy stellar mass until ${\rm log}(M_{\star}/{\rm M_{\odot}})\,\sim\,10.5$
and remains roughly constant for higher masses.
In addition, the metallicity gradients in interacting disc galaxies are
shallower than that in isolated disc galaxies due to effective mixing
\citep[e.g. ][]{Kewley2010, Rupke2010, Torrey2012}.

Indeed, \citet{Pilyugin2014} measured the metallicity of 130
nearby late-type galaxies and derived a common metallicity gradient of
$0.32\,\pm\,0.20\,{\rm dex}\,R_{25}^{-1}$ for 104 of their field galaxies
(i.e. excluding mergers and close pairs). \citet{Sanchez2014} found that
146 galaxies from CALIFA that show no clear evidence of an interaction present
a common metallicity gradient expressed in terms of ${\rm dex}\,R_{\rm e}^{-1}$
with a value $-0.11\,\pm\,0.08\,{\rm dex}\,R_{\rm e}^{-1}$, independent of
morphology, the existence of a bar, absolute magnitude or mass.
\citet{Ho2015} also studied 49 local field isolated spiral
galaxies with absolute magnitudes $-22\,<\,M_{B}\,<\,-16$,
and found evidence for a common metallicity gradient
among their galaxies when the slope is expressed in units of the
isophotal radius $R_{\rm 25}$, i.e. $-0.39\,\pm\,0.18\,{\rm dex}\,R_{25}^{-1}$.
\citet{Pilyugin2019} analysed 147 galaxies from MaNGA and found the metallicity
gradient independent of galaxy stellar with the mean value $-0.2\,{\rm dex}\,R_{25}^{-1}$.
\citet{Bresolin2019} derived the common metallicity gradient with a value
$-0.34\,\pm\,0.2\,{\rm dex}\,R_{\rm 25}^{-1}$ based on long-slit observations
of nearby spiral galaxies.
To compare the best-fitting model predicted observed metallicity gradient with the
observed common metallicity gradient, one must consider the different scale-lengths.
According to the relation between $R_{\rm e}$ and $R_{\rm 25}$, i.e.
$R_{\rm e}\,=\,0.4R_{\rm 25}$ \citep{Zaritsky1994, Sanchez2013}, we can convert the common
metallicity gradient in \citet[][146 isolated spiral galaxies from CALIFA]{Sanchez2014} to
$-0.275\,\pm\,0.2\,{\rm dex}\,R_{\rm 25}^{-1}$.

All the aforementioned information indicates it is worthwhile to investigate
the metallicity gradients for the discs of NGC\,7793, NGC\,300, M33
and NGC\,2403.
Both Fig.\,\ref{Fig:Z_dev} of NGC\,7793 in this work and Fig.\,$2$
of NGC\,2403 in \citet{Kang2017} show that the linear best fit to the observed
data within the optical radius almost coincides with our best-fitting model predicted
metallicity distributions. Here, we also calculate the metallicity gradients of these four
galaxies within their optical radii, and the corresponding values predicted by their
own best-fitting models are separately $-0.492\rm\,dex/R_{25}$ for NGC\,7793,
$-0.384\rm\,dex/R_{25}$ for NGC\,2403, $-0.474\rm\,dex/R_{25}$ for NGC\,300
and $-0.127\rm\,dex/R_{25}$ for M33.
The metallicity gradient of NGC\,7793 agrees well with its observations
\citep{Bibby2010, Pilyugin2014, Stanghellini2015}.
Interestingly, although the models for NGC\,300 in \citet{Kang2016} and
NGC\,2403 in \citet{Kang2017} were constrained by observed metallicities
from HII-region emission line analysis \citep[][NGC\,300]{Bresolin2009}
and \citep[][NGC\,2403]{Berg2013},
their best-fitting model predicted results are in excellent
agreement with the observed metallicity distributions and gradients from
stellar spectral analysis of young massive supergiant stars
\citep[][NGC\,300]{Kudritzki2008, Gazak2015} and \citep[][NGC2403]{Bresolin2022}.
For M33, some new observational metallicity data obtained from
HII-region emission line analysis \citep[e.g. ][]{Alexeeva2022, Rogers2022} and
stellar spectral analysis of young massive blue supergiant stars \citep{Liu2022} were
recently appeared, it would be interesting to consider these metallicities
to constrain the chemical evolution model of M33. Fortunately, our best-fitting
model results of M33 \citep{Kang2012} is basically in agreement with these observations,
taking into account the observed uncertainties and the calibration for determining
metallicity.
It should be emphasized that, although the accurate values of
free parameters in our best-fitting models are not unique, the main
results may not impact our conclusion significantly.

Figure\,\ref{fig:M_gradient} displays metallicity gradient versus stellar
mass. The best-fitting model predictions are shown as black star
(NGC\,7793), red triangle (NGC\,2403), blue square (NGC\,300) and magenta diamond (M33).
Different line types in this Figure represent observed metallicity gradients (expressed
in ${\rm dex}\,R_{\rm 25}^{-1}$) from different authors: dashed lines
\citep{Ho2015}, dotted lines \citep{Sanchez2014} and dash-dotted lines
\citep{Pilyugin2014}. It can be seen that the best-fitting model
predicted metallicity gradients exactly reproduce the observed data.
The metallicity gradients of isolated spiral galaxies NGC\,7793, NGC\,2403
and NGC\,300 are similar and steeper than that of M33, the later is the same
behaviour found by \citet{Rupke2010}, that is, the metallicity gradient
in isolated galaxies is steeper than that in interacting galaxies, since the
interactions between M33 and M31 may change its metallicity gradient.
The similar metallicity gradients in NGC\,7793, NGC\,2403 and NGC\,300
indicate that they may follow very similar chemical evolution when
growing their discs \citep{PB2000, Ho2015}.

%-----------------------------------------------------------------

\section{Conclusions}
\label{sec:sum}

NGC\,7793 is an isolated loosely bound member of the Sculptor Group,
with no bar and a very faint central bulge.
In this work, we build a bridge for the disc of NGC\,7793 between
its observed properties and its evolutionary history by constructing
a simple chemical evolution model. Our results show that the model
predictions are very sensitive to the adopted infall time-scale,
but the outflow process mainly influences the metallicity distributions
along the disc of NGC\,7793, since it takes a fraction of metals away
from NGC\,7793 disc due to its shallower gravitational potential.
The best-fitting model results show that the disc of NGC\,7793
forms inside-out, in excellent agreement with the general predictions
of the inside-out growth scenario for the evolution of spiral galaxies.
About 80\% of the stellar mass of NGC\,7793 is assembled within the last
8\,Gyr and 40\% within the last 4\,Gyr.

We also compared the best-fitting model results of NGC\,7793 with those of
M33, NGC\,300 and NGC\,2403, which are studied in our previous work
\citep{Kang2012, Kang2016, Kang2017}. We found that 78\% (NGC\,7793),
72\% (NGC\,300), 66\% (NGC\,2403) and 79\% (M33) galaxy's stars were
formed between $z\,=\,1$ and $z\,=\,0$, that is, these four galaxies
accumulated more than 50 percent of their total stellar mass within
the past $\sim\rm 8\,Gyr$ (i.e. $z\,=\,1$). Our results also show
that these four disc galaxies simultaneously increase their sizes and
stellar masses as time goes by, and they grow in size at
$\sim\,0.30$ times the rate at which they grow in mass, providing a perfect
match to the previous results \citep{Courteau2007, Pezzulli2015}.
The best-fitting model predicted MZR and gradients, constrained by the
observed metallicities from HII-region emission line analysis, are in excellent
agreement with the metallicities measured from individual massive red and blue
supergiant stars.

\begin{acknowledgements}

We thank the anonymous referee for constructive comments and suggestions,
which improved the quality of our work greatly.
Xiaoyu Kang and Fenghui Zhang are supported by the National Key R\&D Program of China
with (Nos. 2021YFA1600403 and 2021YFA1600400), the National Natural Science Foundation (NSF) of China
(No. 11973081, 11573062, 11403092, 11773063),
the Basic Science Centre project of the NSF of China(No. 12288102),
the Science Research grants from the China Manned Space Project (No. CMS-CSST-2021-A08),
International Centre of Supernovae, Yunnan Key Laboratory (No. 202302AN360001),
the NSF of Yunnan Province (No. 2019FB006).
Rolf Kudritzki acknowledges support by the Munich Excellence Cluster Origins
Funded by the Deutsche Forschungsgemeinschaft (DFG, German Research Foundation)
under the German Excellence Strategy EXC-2094 390783311.

\end{acknowledgements}


\begin{thebibliography}{}

\bibitem[Aller(1942)]{Aller1942} Aller, L.~H.\ 1942, \apj, 95, 52
\bibitem[Alexeeva \& Zhao(2022)]{Alexeeva2022} Alexeeva, S. \& Zhao, G.\ 2022, \apj, 925, 76
\bibitem[Andrews \& Martini(2013)]{Andrews2013} Andrews, B.~H. \& Martini, P.\ 2013, \apj, 765, 140
\bibitem[Asplund et al.(2009)]{Asplund2009} Asplund, M., Grevesse, N., Sauval, A.~J., \& Scott, P.\ 2009, \araa, 47, 481
\bibitem[Aumer \& Binney(2009)]{Aumer2009} Aumer, M. \& Binney, J.~J.\ 2009, \mnras, 397, 1286
\bibitem[Azzollini et al.(2008)]{Azzollini2008} Azzollini, R., Trujillo, I., \& Beckman, J.~E.\ 2008, \apjl, 679, L69
\bibitem[Bakos et al.(2008)]{Bakos2008} Bakos, J., Trujillo, I., \& Pohlen, M.\ 2008, \apjl, 683, L103
\bibitem[Baldry et al.(2008)]{Baldry2008} Baldry, I.~K., Glazebrook, K., \& Driver, S.~P.\ 2008, \mnras, 388, 945
\bibitem[Baldry et al.(2012)]{Baldry2012} Baldry, I.~K., Driver, S.~P., Loveday, J., et al.\ 2012, \mnras, 421, 621
\bibitem[Barnes \& Hernquist(1992)]{Barnes1992} Barnes, J.~E. \& Hernquist, L.\ 1992, \araa, 30, 705
\bibitem[Bauer et al.(2013)]{Bauer2013} Bauer, A.~E., Hopkins, A.~M., Gunawardhana, M., et al.\ 2013, \mnras, 434, 209
\bibitem[Behroozi et al.(2019)]{Behroozi2019} Behroozi, P., Wechsler, R.~H., Hearin, A.~P., et al.\ 2019, \mnras, 488, 3143
\bibitem[Belfiore et al.(2016)]{Belfiore2016} Belfiore, F., Maiolino, R., \& Bothwell, M.\ 2016, \mnras, 455, 1218
\bibitem[Belfiore et al.(2017)]{Belfiore2017} Belfiore, F., Maiolino, R., Tremonti, C., et al.\ 2017, \mnras, 469, 151
\bibitem[Berg et al.(2013)]{Berg2013} Berg, D.~A., Skillman, E.~D., Garnett, D.~R., et al.\ 2013, \apj, 775, 128
%\bibitem[Bergemann et al.(2014)]{Bergemann2014} Bergemann, M., Ruchti, G.~R., Serenelli, A., et al.\ 2014, \aap, 565, A89
\bibitem[Bibby \& Crowther(2010)]{Bibby2010} Bibby, J.~L. \& Crowther, P.~A.\ 2010, \mnras, 405, 2737
\bibitem[Bigiel et al.(2008)]{Bigiel2008} Bigiel, F., Leroy, A., Walter, F., et al.\ 2008, \aj, 136, 2846
\bibitem[Blitz \& Rosolowsky(2006)]{Blitz2006} Blitz, L. \& Rosolowsky, E.\ 2006, \apj, 650, 933
\bibitem[Boissier \& Prantzos(2000)]{BP2000} Boissier, S. \& Prantzos, N.\ 2000, \mnras, 312, 398
\bibitem[Bouquin et al.(2018)]{Bouquin2018} Bouquin, A.~Y.~K., Gil de Paz, A., Mu{\~n}oz-Mateos, J.~C., et al.\ 2018, \apjs, 234, 18
\bibitem[Bresolin et al.(2009)]{Bresolin2009} Bresolin, F., Gieren, W., Kudritzki, R.-P., et al.\ 2009, \apj, 700, 309
\bibitem[Bresolin \& Kennicutt(2015)]{Bresolin2015} Bresolin, F. \& Kennicutt, R.~C.\ 2015, \mnras, 454, 3664
\bibitem[Bresolin(2019)]{Bresolin2019} Bresolin, F.\ 2019, \mnras, 488, 3826
\bibitem[Bresolin et al.(2022)]{Bresolin2022} Bresolin, F., Kudritzki, R.-P., \& Urbaneja, M.~A.\ 2022, \apj, 940, 32
\bibitem[Brinchmann et al.(2004)]{Brinchmann2004} Brinchmann, J., Charlot, S., White, S.~D.~M., et al.\ 2004, \mnras, 351, 1151
\bibitem[Buck(2020)]{Buck2020} Buck, T.\ 2020, \mnras, 491, 5435
\bibitem[Byun \& Freeman(1995)]{Byun1995} Byun, Y.~I. \& Freeman, K.~C.\ 1995, \apj, 448, 563
\bibitem[Calzetti et al.(2015)]{Calzetti2015} Calzetti, D., Lee, J.~C., Sabbi, E., et al.\ 2015, \aj, 149, 51
\bibitem[Casasola et al.(2017)]{Casasola2017} Casasola, V., Cassar{\`a}, L.~P., Bianchi, S., et al.\ 2017, \aap, 605, A18
\bibitem[Chang et al.(1999)]{Chang1999} Chang, R.~X., Hou, J.~L., Shu, C.~G., \& Fu, C.~Q.\ 1999, \aap, 350, 38
\bibitem[Chang et al.(2010)]{Chang2010} Chang, R.~X., Hou, J.~L., Shen, S.~Y., \& Shu, C.~G.\ 2010, \apj, 722, 380
\bibitem[Chang et al.(2012)]{Chang2012} Chang, R.~X., Shen, S.~Y., \& Hou, J.~L.\ 2012, \apjl, 753, L10
%\bibitem[Chang et al.(2017)]{Chang2017} Chang, R., Zhang, S., Shen, S., et al.\ 2017, IAUS, 321, 134.
\bibitem[Chen et al.(2023)]{Chen2023} Chen, B., Hayden, M.~R., Sharma, S., et al.\ 2023, \mnras, 523, 3791
\bibitem[Chiappini et al.(2001)]{Chiappini2001} Chiappini, C., Matteucci, F., \& Romano, D.\ 2001, \apj, 554, 1044
\bibitem[Courteau et al.(2007)]{Courteau2007} Courteau, S., Dutton, A.~A., van den Bosch, F.~C., et al.\ 2007, \apj, 671, 203
%\bibitem[de Jong(1996)]{deJong1996} de Jong, R.~S.\ 1996, \aap, 313, 377
\bibitem[Dalcanton et al.(2009)]{Dalcanton2009} Dalcanton, J.~J., Williams, B.~F., Seth, A.~C., et al.\ 2009, \apjs, 183, 67
%\bibitem[de Blok et al.(2008)]{deBlok2008} de Blok, W.~J.~G., Walter, F., Brinks, E., et al.\ 2008, \aj, 136, 2648
\bibitem[de Vaucouleurs \& Pence(1978)]{Vaucouleurs1978} de Vaucouleurs, G. \& Pence, W.~D.\ 1978, \aj, 83, 1163
\bibitem[de Vaucouleurs et al.(1991)]{1991S&T....82Q.621D} de Vaucouleurs, G., de Vaucouleurs, A., Corwin, H.~G., et al.\ 1991, \skytel, 82, 621
%\bibitem[Dicaire et al.(2008)]{Dicaire2008} Dicaire, I., Carignan, C., Amram, P., et al.\ 2008, \aj, 135, 2038
\bibitem[Edvardsson et al.(1993)]{Edvardsson1993} Edvardsson, B., Andersen, J., Gustafsson, B., et al.\ 1993, \aap, 500, 391
\bibitem[Elmegreen(1989)]{Elmegreen1989} Elmegreen, B.~G.\ 1989, \apj, 338, 178
\bibitem[Erb(2008)]{Erb2008} Erb, D.~K.\ 2008, \apj, 674, 151
\bibitem[Ferguson \& Clarke(2001)]{Ferguson2001} Ferguson, A.~M.~N. \& Clarke, C.~J.\ 2001, \mnras, 325, 781
%\bibitem[Ferreras \& Silk(2001)]{Ferreras2001} Ferreras, I. \& Silk, J.\ 2001, \apj, 557, 165
\bibitem[Feuillet et al.(2019)]{Feuillet2019} Feuillet, D.~K., Frankel, N., Lind, K., et al.\ 2019, \mnras, 489, 1742
\bibitem[Finlator \& Dav{\'e}(2008)]{Finlator2008} Finlator, K. \& Dav{\'e}, R.\ 2008, \mnras, 385, 2181
\bibitem[Frankel et al.(2018)]{Frankel2018} Frankel, N., Rix, H.-W., Ting, Y.-S., et al.\ 2018, \apj, 865, 96
\bibitem[Frankel et al.(2019)]{Frankel2019} Frankel, N., Sanders, J., Rix, H.-W., et al.\ 2019, \apj, 884, 99
\bibitem[Fraternali \& Tomassetti(2012)]{Fraternali2012} Fraternali, F. \& Tomassetti, M.\ 2012, \mnras, 426, 2166
\bibitem[Freedman et al.(2001)]{Freedman2001} Freedman, W.~L., Madore, B.~F., Gibson, B.~K., et al.\ 2001, \apj, 553, 47
%\bibitem[Fu et al.(2009)]{Fu2009} Fu, J., Hou, J.~L., Yin, J., et al.\ 2009, \apj, 696, 668
\bibitem[Gallazzi et al.(2005)]{Gallazzi2005} Gallazzi, A., Charlot, S., Brinchmann, J., et al.\ 2005, \mnras, 362, 41
\bibitem[Garnett et al.(1997)]{Garnett1997} Garnett, D.~R., Shields, G.~A., Skillman, E.~D., et al.\ 1997, \apj, 489, 63
\bibitem[Garnett(2002)]{Garnett2002} Garnett, D.~R.\ 2002, \apj, 581, 1019
\bibitem[Gazak et al.(2015)]{Gazak2015} Gazak, J.~Z., Kudritzki, R., Evans, C., et al.\ 2015, \apj, 805, 182
\bibitem[Goddard et al.(2017)]{Goddard2017} Goddard, D., Thomas, D., Maraston, C., et al.\ 2017, \mnras, 466, 4731
\bibitem[Gogarten et al.(2010)]{Gogarten2010} Gogarten, S.~M., Dalcanton, J.~J., Williams, B.~F., et al.\ 2010, \apj, 712, 858
\bibitem[Gonz{\'a}lez Delgado et al.(2015)]{Gonzalez2015} Gonz{\'a}lez Delgado, R.~M., Garc{\'\i}a-Benito, R., P{\'e}rez, E., et al.\ 2015, \aap, 581, A103
\bibitem[Goswami et al.(2021)]{Goswami2021} Goswami, S., Slemer, A., Marigo, P., et al.\ 2021, \aap, 650, A203
\bibitem[Grisoni et al.(2017)]{Grisoni2017} Grisoni, V., Spitoni, E., Matteucci, F., et al.\ 2017, \mnras, 472, 3637
\bibitem[Grisoni et al.(2018)]{Grisoni2018} Grisoni, V., Spitoni, E., \& Matteucci, F.\ 2018, \mnras, 481, 2570
\bibitem[Guglielmo et al.(2015)]{Guglielmo2015} Guglielmo, V., Poggianti, B.~M., Moretti, A., et al.\ 2015, \mnras, 450, 2749
\bibitem[Haywood(2008)]{Haywood2008} Haywood, M.\ 2008, \mnras, 388, 1175
%\bibitem[Heesen et al.(2014)]{Heesen2014} Heesen, V., Brinks, E., Leroy, A.~K., et al.\ 2014, \aj, 147, 103
\bibitem[Hirschmann et al.(2016)]{Hirschmann2016} Hirschmann, M., De Lucia, G., \& Fontanot, F.\ 2016, \mnras, 461, 1760
\bibitem[Ho et al.(2015)]{Ho2015} Ho, I.-T., Kudritzki, R.-P., Kewley, L.~J., et al.\ 2015, \mnras, 448, 2030
\bibitem[Izotov et al.(2006)]{Izotov2006} Izotov, Y.~I., Stasi{\'n}ska, G., Meynet, G., et al.\ 2006, \aap, 448, 955
\bibitem[Jarrett et al.(2003)]{Jarrett2003} Jarrett, T.~H., Chester, T., Cutri, R., et al.\ 2003, \aj, 125, 525
%\bibitem[Kamphuis \& Briggs(1992)]{Kamphuis1992} Kamphuis, J. \& Briggs, F.\ 1992, \aap, 253, 335
\bibitem[Kang et al.(2012)]{Kang2012} Kang, X., Chang, R., Yin, J., et al.\ 2012, \mnras, 426, 1455
\bibitem[Kang et al.(2015)]{Kang2015} Kang, X., Chang, R., Zhang, F., et al.\ 2015, \mnras, 449, 414
\bibitem[Kang et al.(2016)]{Kang2016} Kang, X., Zhang, F., Chang, R., Wang, L., \& Cheng, L.\ 2016, \aap, 585, A20
\bibitem[Kang et al.(2017)]{Kang2017} Kang, X., Zhang, F., \& Chang, R.\ 2017, \mnras, 469, 1636
\bibitem[Kang et al.(2021)]{Kang2021} Kang, X., Chang, R., Kudritzki, R.-P., et al.\ 2021, \mnras, 502, 1967
\bibitem[Karachentsev et al.(2004)]{Karachentsev2004} Karachentsev, I.~D., Karachentseva, V.~E., Huchtmeier, W.~K., et al.\ 2004, \aj, 127, 2031
\bibitem[Kauffmann(1996)]{Kauffmann1996} Kauffmann, G.\ 1996, \mnras, 281, 475
\bibitem[Kelvin et al.(2014)]{Kelvin2014} Kelvin, L.~S., Driver, S.~P., Robotham, A.~S.~G., et al.\ 2014, \mnras, 444, 1647
%\bibitem[Kennicutt et al.(2003)]{Kennicutt2003} Kennicutt, R.~C., Armus, L., Bendo, G., et al.\ 2003, \pasp, 115, 928
\bibitem[Kennicutt et al.(2011)]{Kennicutt2011} Kennicutt, R.~C., Calzetti, D., Aniano, G., et al.\ 2011, \pasp, 123, 1347
\bibitem[Kewley \& Ellison(2008)]{Kewley2008} Kewley, L.~J. \& Ellison, S.~L.\ 2008, \apj, 681, 1183
\bibitem[Kewley et al.(2010)]{Kewley2010} Kewley, L.~J., Rupke, D., Zahid, H.~J., et al.\ 2010, \apjl, 721, L48
\bibitem[Kobulnicky \& Kewley(2004)]{KK04} Kobulnicky, H.~A. \& Kewley, L.~J.\ 2004, \apj, 617, 240
\bibitem[Kroupa et al.(1993)]{Kroupa1993} Kroupa, P., Tout, C.~A., \& Gilmore, G.\ 1993, \mnras, 262, 545
\bibitem[Kubryk et al.(2013)]{Kubryk2013} Kubryk, M., Prantzos, N., \& Athanassoula, E.\ 2013, \mnras, 436, 1479
\bibitem[Kubryk et al.(2015)]{Kubryk2015} Kubryk, M., Prantzos, N., \& Athanassoula, E.\ 2015, \aap, 580, A126
\bibitem[Kudritzki et al.(2008)]{Kudritzki2008} Kudritzki, R.-P., Urbaneja, M.~A., Bresolin, F., et al.\ 2008, \apj, 681, 269
\bibitem[Kudritzki et al.(2015)]{Kudritzki2015} Kudritzki, R.-P., Ho, I.-T., Schruba, A., et al.\ 2015, \mnras, 450, 342
\bibitem[Kudritzki et al.(2021)]{Kudritzki2021a} Kudritzki, R.-P., Teklu, A.~F., Schulze, F., et al.\ 2021, \apj, 910, 87
%\bibitem[Kudritzki et al.(2021)]{Kudritzki2021b} Kudritzki, R.-P., Teklu, A.~F., Schulze, F., et al.\ 2021b, \apj, 922, 274
\bibitem[Lagos et al.(2011)]{Lagos2011} Lagos, C.~D.~P., Baugh, C.~M., Lacey, C.~G., et al.\ 2011, \mnras, 418, 1649
\bibitem[Larson(1976)]{Larson1976} Larson, R.~B.\ 1976, \mnras, 176, 31
\bibitem[Lee et al.(2011)]{Lee2011} Lee, J.~C., Gil de Paz, A., Kennicutt, R.~C., et al.\ 2011, \apjs, 192, 6
\bibitem[Leitner \& Kravtsov(2011)]{Leitner2011} Leitner, S.~N. \& Kravtsov, A.~V.\ 2011, \apj, 734, 48
%\bibitem[Lequeux et al.(1979)]{Lequeux1979} Lequeux, J., Peimbert, M., Rayo, J.~F., et al.\ 1979, \aap, 80, 155
\bibitem[Leroy et al.(2008)]{Leroy2008} Leroy, A.~K., Walter, F., Brinks, E., et al.\ 2008, \aj, 136, 2782
%\bibitem[Leroy et al.(2009)]{Leroy2009} Leroy, A.~K., Walter, F., Bigiel, F., et al.\ 2009, \aj, 137, 4670
\bibitem[Leroy et al.(2019)]{Leroy2019} Leroy, A.~K., Sandstrom, K.~M., Lang, D., et al.\ 2019, \apjs, 244, 24
\bibitem[Lian et al.(2018a)]{Lian2018a} Lian, J., Thomas, D., \& Maraston, C.\ 2018a, \mnras, 481, 4000
\bibitem[Lian et al.(2018b)]{Lian2018b} Lian, J., Thomas, D., Maraston, C., et al.\ 2018b, \mnras, 476, 3883
\bibitem[Liu et al.(2022)]{Liu2022} Liu, C., Kudritzki, R.-P., Zhao, G., et al.\ 2022, \apj, 932, 29
%\bibitem[Luisi et al.(2018)]{Luisi2018} Luisi, M., Anderson, L.~D., Bania, T.~M., et al.\ 2018, \pasp, 130, 084101
%\bibitem[Lupi et al.(2018)]{Lupi2018} Lupi, A., Bovino, S., Capelo, P.~R., et al.\ 2018, \mnras, 474, 2884
\bibitem[Matteucci \& Francois(1989)]{Matteucci1989} Matteucci, F. \& Francois, P.\ 1989, \mnras, 239, 885
\bibitem[Matteucci(2012)]{Matteucci2012} Matteucci, F.\ 2012, Chemical Evolution of Galaxies: , Astronomy and Astrophysics Library. ISBN 978-3-642-22490-4. Springer-Verlag Berlin Heidelberg, 2012
\bibitem[McCall et al.(1985)]{McCall1985} McCall, M.~L., Rybski, P.~M., \& Shields, G.~A.\ 1985, \apjs, 57, 1
\bibitem[Minchev \& Famaey(2010)]{Minchev2010} Minchev, I. \& Famaey, B.\ 2010, \apj, 722, 112
\bibitem[Minchev et al.(2012)]{Minchev2012} Minchev, I., Famaey, B., Quillen, A.~C., et al.\ 2012, \aap, 548, A126
\bibitem[Moll{\'a} \& D{\'\i}az(2005)]{Molla2005} Moll{\'a}, M. \& D{\'\i}az, A.~I.\ 2005, \mnras, 358, 521
\bibitem[Moustakas et al.(2010)]{Moustakas2010} Moustakas, J., Kennicutt, R.~C., Tremonti, C.~A., et al.\ 2010, \apjs, 190, 233
\bibitem[Mu{\~n}oz-Mateos et al.(2007)]{MM2007} Mu{\~n}oz-Mateos, J.~C., Gil de Paz, A., Boissier, S., et al.\ 2007, \apj, 658, 1006
\bibitem[Mu{\~n}oz-Mateos et al.(2011)]{MM2011} Mu{\~n}oz-Mateos, J.~C., Boissier, S., Gil de Paz, A., et al.\ 2011, \apj, 731, 10
\bibitem[Muraoka et al.(2016)]{Muraoka2016} Muraoka, K., Takeda, M., Yanagitani, K., et al.\ 2016, \pasj, 68, 18
\bibitem[Ostriker et al.(2010)]{Ostriker2010} Ostriker, E.~C., McKee, C.~F., \& Leroy, A.~K.\ 2010, \apj, 721, 975.
\bibitem[Pan et al.(2019)]{Pan2019} Pan, Z., Peng, Y., Zheng, X., et al.\ 2019, \apj, 876, 21
\bibitem[Panter et al.(2008)]{Panter2008} Panter, B., Jimenez, R., Heavens, A.~F., et al.\ 2008, \mnras, 391, 1117
%\bibitem[Peng et al.(2015)]{Peng2015} Peng, Y., Maiolino, R., \& Cochrane, R.\ 2015, \nat, 521, 192
\bibitem[Pezzulli et al.(2015)]{Pezzulli2015} Pezzulli, G., Fraternali, F., Boissier, S., et al.\ 2015, \mnras, 451, 2324
\bibitem[Pilyugin \& Thuan(2005)]{PT05} Pilyugin, L.~S. \& Thuan, T.~X.\ 2005, \apj, 631, 231
\bibitem[Pilyugin et al.(2014)]{Pilyugin2014} Pilyugin, L.~S., Grebel, E.~K., \& Kniazev, A.~Y.\ 2014, \aj, 147, 131
\bibitem[Pilyugin et al.(2019)]{Pilyugin2019} Pilyugin, L.~S., Grebel, E.~K., Zinchenko, I.~A., et al.\ 2019, \aap, 623, A122
\bibitem[Poetrodjojo et al.(2018)]{Poetrodjojo2018} Poetrodjojo, H., Groves, B., Kewley, L.~J., et al.\ 2018, \mnras, 479, 5235
\bibitem[Prantzos \& Boissier(2000)]{PB2000} Prantzos, N. \& Boissier, S.\ 2000, \mnras, 313, 338
\bibitem[Prugniel \& Heraudeau(1998)]{Prugniel1998} Prugniel, P. \& Heraudeau, P.\ 1998, \aaps, 128, 299
\bibitem[Qu et al.(2017)]{Qu2017} Qu, Y., Helly, J.~C., Bower, R.~G., et al.\ 2017, \mnras, 464, 1659
\bibitem[Radburn-Smith et al.(2012)]{Radburn-Smith2012} Radburn-Smith, D.~J., Ro{\v{s}}kar, R., Debattista, V.~P., et al.\ 2012, \apj, 753, 138
\bibitem[Recchi et al.(2008)]{Recchi2008} Recchi, S., Spitoni, E., Matteucci, F., et al.\ 2008, \aap, 489, 555
\bibitem[Rogers et al.(2022)]{Rogers2022} Rogers, N.~S.~J., Skillman, E.~D., Pogge, R.~W., et al.\ 2022, \apj, 939, 44
\bibitem[Romano et al.(2010)]{Romano2010} Romano, D., Karakas, A.~I., Tosi, M., et al.\ 2010, \aap, 522, A32
\bibitem[Ro{\v{s}}kar et al.(2008)]{Rovskar2008} Ro{\v{s}}kar, R., Debattista, V.~P., Stinson, G.~S., et al.\ 2008, \apjl, 675, L65
\bibitem[Rupke et al.(2010)]{Rupke2010} Rupke, D.~S.~N., Kewley, L.~J., \& Chien, L.-H.\ 2010, \apj, 723, 1255
\bibitem[Sacchi et al.(2019)]{Sacchi2019} Sacchi, E., Cignoni, M., Aloisi, A., et al.\ 2019, \apj, 878, 1
\bibitem[Sachdeva et al.(2015)]{Sachdeva2015} Sachdeva, S., Gadotti, D.~A., Saha, K., et al.\ 2015, \mnras, 451, 2
\bibitem[Sachdeva \& Saha(2016)]{Sachdeva2016} Sachdeva, S. \& Saha, K.\ 2016, \apjl, 820, L4
%\bibitem[S{\'a}nchez et al.(2011)]{Sanchez2011} S{\'a}nchez, S.~F., Rosales-Ortega, F.~F., Kennicutt, R.~C., et al.\ 2011, \mnras, 410, 313
\bibitem[S{\'a}nchez et al.(2013)]{Sanchez2013} S{\'a}nchez, S.~F., Rosales-Ortega, F.~F., Jungwiert, B., et al.\ 2013, \aap, 554, A58
\bibitem[S{\'a}nchez et al.(2014)]{Sanchez2014} S{\'a}nchez, S.~F., Rosales-Ortega, F.~F., Iglesias-P{\'a}ramo, J., et al.\ 2014, \aap, 563, A49
\bibitem[Sch{\"o}nrich \& Binney(2009)]{Schonrich2009} Sch{\"o}nrich, R. \& Binney, J.\ 2009, \mnras, 396, 203
%\bibitem[Sch{\"o}nrich \& McMillan(2017)]{Schonrich2017} Sch{\"o}nrich, R. \& McMillan, P.~J.\ 2017, \mnras, 467, 1154
%\bibitem[Schruba et al.(2011)]{Schruba2011} Schruba, A., Leroy, A.~K., Walter, F., et al.\ 2011, \aj, 142, 37
\bibitem[Sellwood \& Wilkinson(1993)]{Sellwood1993} Sellwood, J.~A. \& Wilkinson, A.\ 1993, Reports on Progress in Physics, 56, 173
%\bibitem[Semenov et al.(2019)]{Semenov2019} Semenov, V.~A., Kravtsov, A.~V., \& Gnedin, N.~Y.\ 2019, \apj, 870, 79
\bibitem[Sextl et al.(2023)]{Sextl2023} Sextl, E., Kudritzki, R.-P., Zahid, H.~J., et al.\ 2023, \apj, 949, 60
\bibitem[Skibba et al.(2011)]{Skibba2011} Skibba, R.~A., Engelbracht, C.~W., Dale, D., et al.\ 2011, \apj, 738, 89
\bibitem[Smith et al.(2021)]{Smith2021} Smith, M.~V., van Zee, L., Salim, S., et al.\ 2021, \mnras, 505, 3998
%\bibitem[Spitoni et al.(2010)]{Spitoni2010} Spitoni, E., Calura, F., Matteucci, F., et al.\ 2010, \aap, 514, A73
\bibitem[Spitoni et al.(2015)]{Spitoni2015} Spitoni, E., Romano, D., Matteucci, F., et al.\ 2015, \apj, 802, 129
%\bibitem[Spitoni et al.(2017)]{Spitoni2017} Spitoni, E., Vincenzo, F., \& Matteucci, F.\ 2017, \aap, 599, A6
\bibitem[Spitoni et al.(2019)]{Spitoni2019} Spitoni, E., Cescutti, G., Minchev, I., et al.\ 2019, \aap, 628, A38
\bibitem[Spitoni et al.(2020)]{Spitoni2020} Spitoni, E., Calura, F., Mignoli, M., et al.\ 2020, \aap, 642, A113
\bibitem[Spitoni et al.(2021a)]{Spitoni2021a} Spitoni, E., Verma, K., Silva Aguirre, V., et al.\ 2021a, \aap, 647, A73
\bibitem[Spitoni et al.(2021b)]{Spitoni2021b} Spitoni, E., Calura, F., Silva Aguirre, V., et al.\ 2021b, \aap, 648, L5
\bibitem[Stanghellini et al.(2015)]{Stanghellini2015} Stanghellini, L., Magrini, L., \& Casasola, V.\ 2015, \apj, 812, 39
%\bibitem[Stark et al.(1987)]{Stark1987} Stark, A.~A., Elmegreen, B.~G., \& Chance, D.\ 1987, \apj, 322, 64
%\bibitem[Tamburro et al.(2008)]{Tamburro2008} Tamburro, D., Rix, H.-W., Walter, F., et al.\ 2008, \aj, 136, 2872
\bibitem[Thomas et al.(2010)]{Thomas2010} Thomas, D., Maraston, C., Schawinski, K., et al.\ 2010, \mnras, 404, 1775
\bibitem[Tinsley(1980)]{Tinsley1980} Tinsley, B.~M.\ 1980, \fcp, 5, 287
\bibitem[Torrey et al.(2012)]{Torrey2012} Torrey, P., Cox, T.~J., Kewley, L., et al.\ 2012, \apj, 746, 108
\bibitem[Tremonti et al.(2004)]{Tremonti2004} Tremonti, C.~A., Heckman, T.~M., Kauffmann, G., et al.\ 2004, \apj, 613, 898
%\bibitem[Trussler et al.(2020)]{Trussler2020} Trussler, J., Maiolino, R., Maraston, C., et al.\ 2020, \mnras, 491, 5406
\bibitem[van den Bosch(1998)]{vandenBosch1998} van den Bosch, F.~C.\ 1998, \apj, 507, 601
\bibitem[Vincenzo et al.(2016)]{Vincenzo2016} Vincenzo, F., Matteucci, F., Belfiore, F., et al.\ 2016, \mnras, 455, 4183
\bibitem[Vincenzo \& Kobayashi(2020)]{Vincenzo2020} Vincenzo, F. \& Kobayashi, C.\ 2020, \mnras, 496, 80
\bibitem[Walter et al.(2008)]{Walter2008} Walter, F., Brinks, E., de Blok, W.~J.~G., et al.\ 2008, \aj, 136, 2563
\bibitem[Williams et al.(2009)]{Williams2009} Williams, B.~F., Dalcanton, J.~J., Dolphin, A.~E., et al.\ 2009, \apjl, 695, L15
\bibitem[Williams et al.(2011)]{Williams2011} Williams, B.~F., Dalcanton, J.~J., Johnson, L.~C., et al.\ 2011, \apjl, 734, L22
\bibitem[Williams et al.(2013)]{Williams2013} Williams, B.~F., Dalcanton, J.~J., Stilp, A., et al.\ 2013, \apj, 765, 120
\bibitem[Wolfe et al.(2013)]{Wolfe2013} Wolfe, S.~A., Pisano, D.~J., Lockman, F.~J., et al.\ 2013, \nat, 497, 224
\bibitem[Xiang \& Rix(2022)]{Xiang2022} Xiang, M. \& Rix, H.-W.\ 2022, \nat, 603, 599
%\bibitem[York et al.(2000)]{York2000} York, D.~G., Adelman, J., Anderson, J.~E., et al.\ 2000, \aj, 120, 1579.
%\bibitem[Yin et al.(2009)]{Yin2009} Yin, J., Hou, J.~L., Prantzos, N., et al.\ 2009, \aap, 505, 497
\bibitem[Zahid et al.(2014)]{Zahid2014} Zahid, H.~J., Dima, G.~I., Kudritzki, R.-P., et al.\ 2014, \apj, 791, 130
\bibitem[Zahid et al.(2017)]{Zahid2017} Zahid, H.~J., Kudritzki, R.-P., Conroy, C., et al.\ 2017, \apj, 847, 18
\bibitem[Zaritsky et al.(1994)]{Zaritsky1994} Zaritsky, D., Kennicutt, R.~C., \& Huchra, J.~P.\ 1994, \apj, 420, 87
\bibitem[Zheng et al.(2017)]{Zheng2017} Zheng, Z., Wang, H., Ge, J., et al.\ 2017, \mnras, 465, 4572


\end{thebibliography}

\end{document}